\documentclass[pra,reprint,showpacs,showkeys,groupedaddress,nofootinbib]{revtex4-1}
\usepackage{amsmath,amssymb,amsfonts}
\usepackage{graphicx}
\usepackage{txfonts}
\usepackage{epstopdf}
\usepackage[T1]{fontenc}

\usepackage[unicode]{hyperref}
\hypersetup{
   unicode=true,          
   a4paper=true,
   plainpages=false,
   pdftitle={Title of PDF},    
   pdfauthor={Author of PDF},     
   pdfsubject={Subject of PDF},   
   colorlinks=true,       
}
\newcommand{\bra}[1]{\langle #1\vert}
\newcommand{\ket}[1]{\vert #1\rangle}
\newcommand{\bket}[2]{\langle #1\vert#2\rangle}
\newcommand{\ev}[1]{\langle #1 \rangle}
\newcommand{\pd}[1]{\frac{\partial}{\partial #1}}
\newcommand{\pdsq}[1]{\frac{\partial^2}{\partial #1^2}}
\newcommand{\refeq}[1]{Eq.~(\ref{#1})}

\newcommand{\reffig}[1]{Fig.~\ref{#1}}
\newcommand{\sech}[0]{\mbox{sech}}
\newcommand{\scut}[0]{\eta}

\begin{document}

\title{Quantum theory of bright matter wave solitons in harmonic confinement}
\author{David I. H. Holdaway}
\email{d.i.h.holdaway@dur.ac.uk}
\author{Christoph Weiss}

\author{Simon A. Gardiner}

  \affiliation{Department of Physics, Durham University, Durham DH1 3LE, United Kingdom}

\date{10 May 2012}

 \begin{abstract}
This paper investigates bright quantum-matter-wave solitons beyond the Gross Pitaevskii equation (GPE).  As proposals for interferometry and creating nonlocal quantum superpositions have been formed, it has become necessary to investigate effects not present in mean field models.  We investigate the effect of harmonic confinement on the internal degrees of freedom, as the ratio of zero point harmonic oscillator length to classical soliton length, for different numbers of atoms.  We derive a first-order energy correction for the addition of a harmonic potential to the many-body wavefunction and use this to create a variational technique based on energy minimization of this wavefunction for an arbitrary number of atoms, and include numerics based on diagonalization of the Hamiltonian in a basis of harmonic oscillator Fock states.  Finally we compare agreement between a Hartree product ground state and the Bethe ansatz solution with a Gaussian envelope localizing the center of mass and show a region of good agreement.  

\end{abstract}


\maketitle

\section{Introduction}

Since the experimental realization of Bose-Einstein condensation (BEC) with dilute atomic gases~\cite{AndersonEnsher1995,DavisMewes1995}, much progress has been made in the degree of control possible in terms of external potentials and control over interactions via magnetic and optical Feshbach resonances.  Recently, it has become possible to create condensates of atomic species with scattering lengths that can be tuned to be negative~\cite{KhaykovichSchreck2002,StreckerPartridge2002,CornishThompson2006}, for example hyperfine levels of $^{85}$Rb, $^{7}$Li  and $^{133}$Cs~\cite{ChinVuletic2004,VogelsTsai1997}. Chromium is also shown to have negatively tunable scattering lengths, but also has significant long-range dipole forces~\cite{BeaufilsChicireanu2008}.  

Such attractive condensates have the remarkable property of `self trapping', i.e.\ being localized (at least in terms of pair correlations) on a length scale shorter than would be expected for a non-interacting condensate.  In the limit in which there is one dimension where the non-interacting ground state would be infinitely wide, the Gross Pitaevskii equation (GPE) predicts the ground state would still be localized to a finite size.
Therefore with the GPE, these condensates behave as solitary matter waves, or in the quasi 1D case, as classical solitons.  The transitional region between 1D and 3D  has been investigated using variational methods~\cite{SalasnichParolaReatto2002}.  Experiments measured systems with a lifetime of several seconds, both for the case of a single wavepacket~\cite{KhaykovichSchreck2002} (a ground state) and multiple smaller wavepackets~\cite{StreckerPartridge2002,AlKhawajaEtAl02}, referred to as soliton trains.  In addition to experimental results, theoretical results relating to the interaction of such systems with potential barriers predicted effects such as enhanced reflection and transmission~\cite{GarnierAbdullaev2006,LeeBrand2006}.

In this situation, the GPE predicts an infinite number of conserved quantities within the system and thus complete integrability~\cite{ZhaoLuo2008}. As a result the inverse scatting transform can be used to obtain solutions~\cite{AblowitzSegurBook1981} that are a combination of bright solitons, which do not change shape as the quantum pressure is exactly balanced by the nonlinear interaction, and radiation, which does.  In fact any initial condition for the GPE equation can be broken up into these components~\cite{AblowitzSegurBook1981}. In the case of systems of a multiple soliton system, individual solitons can collide with other solitons without a transfer of energy between them, resulting in only an asymptotic position and phase shift. 
Such localized matter waves (which are typically of the order of a few micrometers in width) could also theoretically be split coherently into multiple parts~\cite{BillamCornish2011} and prove useful for interferometry~\cite{MartinRuostekoski2011,HelmEtAl2012}, studies of quantum reflection~\cite{LeeBrand2006,CornishParker2009} or probing surface potentials~\cite{McGuirkHarber2004}.  

Under certain circumstances such solitons behave like classical particles with finite range interactions~\cite{MartinAdams2008} even in the presence of harmonic confinement. However, in reality, such objects should behave as quantum particles (i.e.\ with no substructure, but with the location determined by a wave-function obeying quantum mechanical laws rather than a specific position). Despite the GPEs success in describing many phenomena in BEC, even for very small numbers of atoms~\cite{MazetsKurizki2006}, this quantum mechanical center-of-mass behavior is totally lost under the approximation of a product state wave-function.  Therefore we consider a full many-body quantum description, making use of the usual pseudo-potential approximation.

The dynamics of the center of mass of an interacting gas in a harmonic potential are independent of the interactions, giving rise to the so called ``Kohn mode'' ~\cite{BonitzBalzer2007}.  More generally, any potential which looks locally harmonic on the length scales dictated by the internal degrees of freedom (in this case, the classical soliton length) can be considered to only weakly couple the center-of-mass to other degrees of freedom.  As a result, this behavior needs to be considered separately. In this weak coupling approximation the center-of-mass behaves like a non-interacting particle of mass $N M$, which can be localized on scales far wider than a classical soliton length or even not at all.  This delocalisation is not present in the mean field model, however the Kohn mode is still present for translational motion of the center of mass.  

Exact results exist for Bose gases in free space with periodic boundary conditions for both repulsive~\cite{LiebLiniger1963} and attractive ~\cite{McGuire1964,CalabreseCaux2007} interactions; these have the advantage of explicitly separating the center-of-mass component of the wave-function and being accurate even for very small numbers of atoms and fragmented states, for which the GPE is not. It has however been shown that in certain situations, in the limit $N \to \infty$, $g \to 0$ with $N g = {\rm constant}$, the GPE functional is an exact description of the system~\cite{LiebSeiringer2000}.  Modern numerical approaches are available, such as the multi-configurational-time-dependent Hartree method, which have been used to study bright matter wave solitons~\cite{StreltsovAlon2009}. 
The separation properties of the center-of-mass in free space and harmonic traps make available the possibilities of creating non local superpositions~\cite{WeissCastin2009,StreltsovAlon2009} (cf.~\cite{LewensteinMalomed2009}) with sufficient numbers of atoms to make them detectable, in ways not predicted in the classical field description of the GPE.

In the following we study the Lieb-Liniger(-McGuire) gas~\cite{LiebLiniger1963,McGuire1964}, a 1D system of identical Bosons with attractive contact interactions, with the addition of a harmonic trapping potential. We have present a series of analytic and numerical techniques to study many-body effects, making use of the separability of the many-body wave-function into a center-of-mass component and a relative component.  This paper focuses on the ground state of the system and energy corrections to the relative components of the wave-function as trapping is increased, along with estimates of the overlap with the free-space relative ground state.  

We consider a unit system with the Hartree soliton length set to unity and $\sqrt{\gamma}$ a dimensionless ratio between this length and the harmonic oscillator length in the axial direction, such that $\gamma \to 0$ recovers the free case solved by the Bethe Ansatz.  We use a numerical method based on exact diagonalization of the Hamiltonian over a basis set of Hermite functions, truncated up to a maximum energy, in order to determine the many body ground state energies, combined with a variational method for low $\gamma$.  

This paper is organized as follows: 
Section~\ref{sec:krs} introduces the exact results in one dimensional infinite systems (using the Lieb-Liniger model~\cite{LiebLiniger1963}) and the unit rescaling used to keep the mean field soliton length constant throughout the paper.  Also included is the separability of the many-body Hamiltonian and the existence of the Kohn mode as well as the exact eigenstates for two interacting bosons in a harmonic potential.
Section~\ref{sec:pert and var} derives a perturbative energy correction to the relative ground state energy from the introduction of a harmonic trapping potential, along with a variational procedure to estimate the ground state in the limit of weak trapping.  
Section~\ref{sec:hop} introduces the numerical method used to perform calculations in the many-body system for varying 1D harmonic trapping potential, using a basis set of harmonic oscillator eigenstates, which are projected to a center-of-mass excitation basis.   
Section~\ref{sec:results} numerically investigates changes to relative component (i.e. having excluded the center of mass) of the ground state as the trapping potential is increased. These calculations are performed for different numbers of atoms and compared with predictions based on the GPE. Section~\ref{sec:agreement} examines quantitatively the overlap between the mean field approximation of the ground state and the many-body solutions, along with finding a regime of agreement where the difference between the two models is small in every respect.  
Section~\ref{sec:conc} summarizes and comments on the results.

\section{Preliminaries \label{sec:krs}}

\subsection{System overview and rescaling}

\subsubsection{The many-body Hamiltonian in first and second quantization}
We consider a system of identical bosonic atoms within a cylindrically symmetric prolate (the radial frequency $\omega_{r}$ is greater than the axial frequency $\omega_{x}$) harmonic trapping potential $V(x,y,z) = M[\omega_{x}^{2}x^{2}+\omega_{r}^{2}(y^{2}+z^{2})]/2$, where $M$ is the atomic mass.  We further consider the system to be sufficiently low-temperature for the atomic interactions to be pure $s$-wave and contact-like, and assume that we are in an appropriate parameter regime so that the radial modes can be considered ``frozen out'' for low-energy states, taking the Gaussian form of radial harmonic oscillator ground states \cite{BillamWrathmall2011b,ParkerCornish2007,Olshanii1998}.  Finally, we assume the interactions to be attractive.

Integrating out the radial degrees of freedom, we obtain the 1D Hamiltonian, in second-quantized form, and hence for an arbitrary number of atoms, as follows:
\begin{multline}
 \hat{H} =  \int  dx  \hat{\Psi}^{\dagger}(x)\left[-\frac{\hbar^2}{2M}\pdsq{x}+\frac{M\omega_{x}^2 x^2}{2}
 \right]\hat{\Psi}(x) 
 \\
- \frac{g_{\textrm{1D}}}{2}  \int  dx \hat{\Psi}^{\dagger}(x)\hat{\Psi}^{\dagger}(x)\hat{\Psi}(x)\hat{\Psi}(x),
\label{Eq:SecondQuantizedH}
\end{multline}
where $g_{\textrm{1D}} = 2\hbar\omega_{r}\vert a_{s}\vert $, and $a_{s}$ is the (assumed negative) $s$-wave scattering length~\cite{PitaevskiiStringariBook2003}.  The coordinate-space representation for the corresponding first-quantized form of this Hamiltonian for $N$ atoms is then
\begin{equation}
H(\vec{x}) = \sum_{k=1}^N \left( -\frac{\hbar^{2}}{2M} \pdsq{x_{k}} + \frac{M\omega_x^2 x_{k}^2 }{2} \right)
 - g_{\textrm{1D}} \sum_{k=2}^{N} \sum_{j =1}^{k-1} \delta(x_{k}-x_{j}),
\label{Eq:FirstQuantizedH}
\end{equation}
where we use $\vec{x}$ as a shorthand for $\{x_{1},x_{2},x_{3},\ldots,x_{N}\}$, the $N$ individual particle coordinates.  In the absence of any trapping potential (i.e., $\omega_x=0$), the system is formally integrable \cite{LiebLiniger1963}, and the first exact results for the case of attractive interactions were obtained in 1964 by McGuire \cite{McGuire1964}. 

\subsubsection{Hartree factorization: The Gross-Pitaevskii equation}

In this approximation one assumes the many-body wavefunction $\psi(\vec{x})$ to be factorizable into product form, such that each individual atom is described by the same single-particle
wavefunction $\phi(x_{k})$. Hence, the Hartree wavefunction $\psi_{\textrm{H}}(\vec{x}) = \prod_{k=1}^N \phi(x_{k})$.
Minimizing the energy $E = \int d\vec{x} \psi_{\rm H}^{\star}(\vec{x}) H(\vec{x}) \psi_{\rm H}(\vec{x})$ of such a stationary state with respect to variations in  $\phi(x_{k})$ leads to~\cite{LaiHaus1989A}
\begin{equation}
\mu\phi(x) = \left[ -\frac{\hbar^2}{2M} \pdsq{x} + \frac{M\omega_x^2 x^2}{2} 
   - g_{\textrm{1D}}(N-1)  \vert \phi(x) \vert^2 \right]\phi(x),
\label{Eq:GPE}
\end{equation}
with $\phi(x)$ normalized to unity, and $\mu$ a Lagrange multiplier, given by
\begin{multline}
\mu = \int dx \phi^{*}(x)\left[
-\frac{\hbar^2}{2M} \pdsq{x} + \frac{M\omega_x^2 x^2}{2} 
\right]\phi(x)
\\- g_{\textrm{1D}}(N-1)  \int dx  \vert \phi(x) \vert^{4}. 
\label{Eq:GPEEigenvalue}
\end{multline}
Equation (\ref{Eq:GPE}) is the one-dimensional time-independent Gross-Pitaevskii equation (GPE) \cite{PitaevskiiStringariBook2003}, which formally tends to an exact description as $N\rightarrow \infty$ while $g_{\textrm{1D}}N$ is held constant \cite{CastinDum1998,GardinerMorgan2007}.  In this limit $g_{\textrm{1D}}(N-1) \approx g_{\textrm{1D}}N$, and it is more typical for the coefficient of the nonlinearity in Eq.\ \eqref{Eq:GPE} to be set proportional to $N$.  As we will also consider small particle numbers, in what follows we choose to retain the proportionality to $(N-1)$.

\subsubsection{Rescaling to dimensionless form \label{sec:rescale}}

It is convenient to rescale our description of the system in terms of an effective $\hbar=M=g_{\textrm{1D}}(N-1) = 1$ unit system, referred to as ``soliton units'' \cite{MartinAdams2008,BillamWrathmall2011b}. Space, time and energy scales are then given in units of
$\hbar^{2}/Mg_{\textrm{1D}}(N-1)$ (the classical soliton length \cite{BillamWrathmall2011b}), $\hbar^3/Mg_{\textrm{1D}}^{2}(N-1)^{2}$, and $Mg_{\textrm{1D}}^{2}(N-1)^{2}/\hbar^{2}$, respectively.  

We work within this system of units from this point onwards.  Equation \eqref{Eq:SecondQuantizedH} then simplifies to 
\begin{multline}
 \hat{H} =  \int dx \hat{\Psi}^{\dagger}(x)\left[-\frac{1}{2}
 \pdsq{x}+\frac{\gamma^{2} x^2}{2}\right]\hat{\Psi}(x)
\\- \frac{1}{2(N-1)}  \int dx \hat{\Psi}^{\dagger}(x)\hat{\Psi}^{\dagger}(x)\hat{\Psi}(x)\hat{\Psi}(x),
\label{Eq:ScaledSecondQuantizedH}
\end{multline}
the first-quantized form of the Hamiltonian [\refeq{Eq:FirstQuantizedH}] transforms to
\begin{equation}
H(\vec{x}) = \sum_{k=1}^N \left[-\frac{1}{2} \pdsq{x_{k}} + \frac{\gamma^{2} x_{k}^2 }{2} \right]
-  \frac{1}{N-1}\sum_{k=2}^{N} \sum_{j =1}^{k-1} \delta ( x_{k}-x_{j}),
\label{Eq:ScaledFirstQuantizedH}
\end{equation}
and the GPE [\refeq{Eq:GPE}] becomes 
\begin{equation}
 \mu\phi(x) = \left[-\frac{1}{2} \pdsq{x} + \frac{\gamma^{2}x^{2}}{2}- \vert \phi(x) \vert^2 \right] \phi(x).
\label{Eq:ScaledGPE}
\end{equation}

We have introduced the dimensionless parameter $\gamma$, which is the square of the ratio of the classical soliton length to the harmonic length $\sqrt{\hbar/M\omega_{x}}$ \cite{BillamWrathmall2011b}, i.e.,
\begin{equation}
\label{eq:gammadef}
\gamma = \frac{\hbar^3\omega_x}{Mg_{\textrm{1D}}^{2}(N-1)^{2}}.
\end{equation}
Within our chosen system of units $\gamma$ appears in the rescaled Hamiltonian and GPE as a dimensionless effective trap frequency.  This also reveals $\gamma$ to be the only free parameter in the GPE, which as a description of the system is effectively a classical field limit, and the particle number $N$ appears as an additional free parameter in the fully quantal Hamiltonian.

\subsection{Known exact results}
\subsubsection{Exact results in free space}
In the case where there is no axial trapping potential, i.e., $\gamma = 0$, the many-body eigenstates~\cite{McGuire1964,LaiHaus1989B}
and GPE stationary states~\cite{AblowitzSegurBook1981,LaiHaus1989A,BillamWrathmall2011b} 
are known.  The stationary solutions to the GPE [\refeq{Eq:ScaledGPE}] minimizing the energy are classical bright solitons \cite{MartinAdams2008,BillamWrathmall2011b}
\begin{equation}
 \phi(x) = \frac{1}{2}\sech\left(\frac{x-x_0}{2}\right),
\label{Eq:ClassicalSoliton}
\end{equation}
where the value of $x_{0}$ is arbitrary (as is that of an irrelevant global phase). The Hartree approximation to the many-body wavefunction is thus
\begin{equation}
 \psi_{\textrm{H}}(\vec{x}) = \frac{1}{2^{N}} \prod_{k=1}^{N} \sech\left(\frac{x_{k}-x_0}{2}\right),
\label{Eq:HartreeSolution}
\end{equation}
and is localized around $x_{0}$, i.e., $\psi_{\textrm{H}}(\vec{x})\to 0$ as $\vert x_{k}-x_0\vert \to \infty$.  Exact solutions also exist in box and periodic boundary conditions, given by Jacobi elliptic functions~\cite{CarrClark2000}.

The exact ground state wavefunction for \refeq{Eq:ScaledFirstQuantizedH} with $\gamma = 0$ (see Appendices~\ref{betheans} and \ref{sec:norm}) is an $N$-particle bound state, proportional to~\cite{McGuire1964,LaiHaus1989B}
\begin{equation}
\psi_{\textrm{G}}(\vec{x}) = 
\sqrt{\frac{(N-1)!}{(N-1)^{N-1}}}
\exp\left(-\sum_{k=2}^{N} \sum_{j =1}^{k-1}  \frac{\vert x_{k}-x_{j} \vert}{2[N-1]}  \right).
\label{Eq:FreeBoundState}
\end{equation}
As the Hamiltonian [\refeq{Eq:ScaledFirstQuantizedH}] has no external potential, all its eigenfunctions are independent of the center-of-mass coordinate 
\begin{equation}
x_{\textrm{C}} = \frac{1}{N}\sum_{k=1}^{N}x_{k},
\label{Eq:CoMDefinition}
\end{equation}
(see section \ref{Sec:Separation}) and there exists a continuum of moving $N$-particle bound state eigenfunctions $ \psi(P,\vec{x})=e^{i P x_{\rm C}} \psi_{\textrm{G}}(\vec{x})/\sqrt{2 \pi}$.  The normalization convention is then such that $\int d\vec{x} \psi^{*}(P,\vec{x})\psi(P',\vec{x}) = \delta(P-P')$, and the ground state is written as $\psi(0,\vec{x}) \equiv \psi_{\textrm{G}}(\vec{x})/\sqrt{2 \pi}$.  The exact ground state is thus completely delocalized in the center-of-mass,  and is only localized in the sense that $\psi_{\textrm{G}}(\vec{x})\rightarrow 0$ as $\vert x_{k}-x_{j} \vert \rightarrow \infty$, for any $k,j$.  

The localized Hartree solution $\psi_{\textrm{H}}(\vec{x})$ violates the translational symmetry requirement imposed by the absence of external potentials in the Hamiltonian due to the tacit assumption that the minimizing wavefunction should vanish as $x_{k}\to\pm\infty$ \cite{LaiHaus1989A}.  Note, however, that the particle densities about a specified value $R$ of the  center-of-mass coordinate [given by the expectation value of $\delta(R - x_{\textrm{C}}) \sum_{k=1}^N \delta(x-x_{k})$] corresponding to $\psi_{\textrm{H}}(\vec{x})$ [\refeq{Eq:HartreeSolution}] and $ \psi_{\textrm{G}}(\vec{x})$ [\refeq{Eq:FreeBoundState}] agree to order $1/N$~\cite{CalogeroDegasperis1975}, and hence are identical in the limit $N\to \infty$.  The energies $E_{\textrm{H}}$ and $E_{\textrm{G}}$ corresponding to the wavefunctions $ \psi_{\textrm{H}}(\vec{x})$ and $\psi_{\textrm{G}}(\vec{x})$ are given by
\begin{align}
E_{\textrm{H}}&= -\frac{N}{24},  
\label{Eq:FreeHartreeEnergy}
\\
E_{\textrm{G}} &= -\frac{N(N+1)}{24(N-1)}
\equiv E_{\textrm{H}} - \frac{N}{12(N-1)}.
\label{Eq:FreeFullEnergy}
\end{align}

As one would expect, the exact eigenenergy $E_{\textrm{G}}$ is less than $E_{\textrm{H}}$, and the
difference in energies \textit{per particle\/} $(E_{\textrm{H}}-E_{\textrm{G}})/N=1/12(N-1)$ vanishes as $N\rightarrow\infty$.

\subsubsection{Separation of the center-of-mass coordinate $x_{\rm C}$ \label{Sec:Separation}}
In the case of any external potential being either harmonic or nonexistent, the center-of-mass dynamics separate and are independent of any two-body interactions. Consequently the center-of-mass eigenstates are simple harmonic oscillator eigenstates or plane waves, respectively, in the former case this is referred to as the Kohn mode.  This may be readily seen by expressing the first-quantized form of the Hamiltonian \refeq{Eq:ScaledFirstQuantizedH} in terms of the
Jacobi coordinates, i.e., $x_{\textrm{C}}$ [\refeq{Eq:CoMDefinition}] together with
\begin{equation}
\xi_{k} \equiv  x_{k} - \frac{1}{k-1}\sum_{j=1}^{k-1} x_{j},
\label{Eq:JacobiCoordinates}
\end{equation}
for $k\in \{2,3,4,\ldots,N\}$.  The Hamiltonian can then be phrased as $H = H_{\textrm{C}} + H_{\textrm{R}}$, where
\begin{align}
H_{\textrm{C}}(x_{\textrm{C}}) =  &-\frac{1}{2N} \pdsq{x_{\textrm{C}}} + \frac{N \gamma^{2}x_{\textrm{C}}^2}{2},  
\label{Eq:HamiltonianC}
\\
\begin{split}
H_{\textrm{R}}(\vec{\xi}) = &\sum_{k=2}^N \left[ - \frac{k}{2(k-1)} \pdsq{\xi_k} + \frac{ (k-1) \gamma^{2}\xi_k^2 }{2k} 
\right]  
\\
&- \frac{1}{N-1} \sum_{k=2}^{N}\delta\left(\xi_k + \sum_{\ell=2}^{k-1} \frac{\xi_\ell}{\ell} \right)
\\
&- \frac{1}{N-1} \sum_{k=2}^{N}\sum_{j=2}^{k-1} \delta\left(\xi_k + \sum_{\ell=j+1}^{k-1} \frac{\xi_\ell}{\ell} -  \frac{j-1}{j}\xi_j \right),
\end{split}
\label{Eq:HamiltonianR}
\end{align}
$\vec{\xi}$ is a shorthand for $\{\xi_{2},\xi_{3},\xi_{4},\ldots,\xi_{N}\}$, and we have used the identity 
$x_{k} - x_{j} = \xi_{k} + \sum_{\ell=j+1}^{k-1}\xi_{\ell}/\ell - [(j-1)/j]\xi_{j}$ (with 
$b>a$ and $\xi_{1}\equiv x_{\textrm{C}}$).  In cases where the upper limit of a sum is less than its lower limit, the sum is taken $=0$.  

Hence, the normalized ground state of $H_{\textrm{C}}$ is exactly
\begin{equation}
\psi_{\textrm{C}} (x_{\textrm{C}})= 
\left(
\frac{N\gamma}{\pi}
\right)^{1/4}
\exp
\left(
-\frac{N\gamma x_{\textrm{C}}^{2}}{2}
\right),
\label{Eq:CoMGround}
\end{equation}
with eigenenergy $=\gamma/2$.

\subsubsection{Two interacting bosons in a harmonic potential \label{sec:2 bosons}}
The case of two identical bosons in a harmonic potential with contact ($\delta$-function) interactions is also exactly solvable \cite{BuschEnglert1998,SowinskiBrewczyk2010}. In this case the eigenfunctions of $H_{\textrm{R}}(\xi_{2})$, defined through $H_{\textrm{R}}(\xi_{2})\phi_{n}(\xi_2) = E_{\textrm{R},n}\phi_{n}(\xi_2)$, are given by
\begin{equation}
\phi_{n}(\xi_2) =\mathcal{N}_{n} U(-\nu_n,1/2,\gamma\xi_{2}^2/2)e^{-\gamma \xi_2^2/4},
\end{equation}
where  $U(a,b,z)$ is the Tricomi confluent hypergeometric function \cite{WolframTricomiFunction}, and $\mathcal{N}_{n}$ is a normalization constant.  The $\nu_{n}$ are implicit solutions of 
\begin{equation} 
\frac{\Gamma(1/2-\nu_n)}{\Gamma(-\nu_n)} = \frac{1}{2\sqrt{2 \gamma}},
\label{Eq:GammaFunctionIdentity}
\end{equation}
and set the eigenvalues of $H_{\textrm{R}}(\xi_{2})$ through
\begin{equation}
 E_{\textrm{R},n} = \left(2\nu_n+\frac{1}{2}\right)\gamma.
\label{Eq:TwoBodyEnergy}
\end{equation}
Attractive interactions must reduce $E_{\textrm{R},0}$ from the noninteracting case, so that $E_{\textrm{R},0} < \gamma/2 \Rightarrow \nu_{0} < 0$.  As outlined in appendix \ref{App:GroundStateEnergy}, it then follows that in the limit $\gamma \to 0$ (interaction dominated regime) $E_{\textrm{R},0} \to -1/4 + \mathcal{O}(\gamma^{2})$.  This is in agreement with the \textit{total\/} ground state energy $E_{\textrm{G}}$ for the case of two attractively interacting bosons in free space [\refeq{Eq:FreeFullEnergy}], as one would expect due to the  center-of-mass energy of the free space ground state being $=0$.
In the opposite limit of $\gamma^{-1}\to 0$ (trap dominated regime) harmonic oscillator eigenvalues and eigenfunctions must result, i.e., $E_{n}\to (2n+1/2)\gamma$ and $U(-\nu_{n},1/2,\gamma\xi_{2}^{2}/2)\rightarrow H_{2n}(\sqrt{\gamma}\xi_{2})/2^{2n}/\sqrt{2}$, where the $H_{2n}$ are even Hermite polynomials.\footnote{Due to Bose symmetry $\phi_{n}(\xi_{2}) \equiv \phi_{n}(-\xi_{2})$, i.e., eigenfunctions must be even.}

\section{Perturbative and variational methods\label{sec:pert and var}}

\subsection{Interaction dominated limit in a harmonic potential}

In the case where $\gamma \ll 1$, we may consider the effect of the trap to be dominated by the effect of the interactions, and therefore negligible in $H_{\textrm{R}}$. As there are no interactions present in $H_{\textrm{C}}$,  the effect of the trap is in this case always significant, even in the interaction dominated regime.  

We may therefore consider a limiting case Hamiltonian $H_{0}$, composed of $H_{\textrm{R}}$ [\refeq{Eq:HamiltonianR}] with $\gamma = 0$, plus $H_{\textrm{C}}$ [\refeq{Eq:HamiltonianC}].  Written in terms of conventional single particle coordinates,
\begin{equation}
H_{0}(\vec{x}) = -\frac{1}{2} \sum_{k=1}^N  \pdsq{x_{k}} 
+ \frac{\gamma^2}{2N} \left(\sum_{k=1}^N  x_{k}\right)^2 
- \frac{1}{N-1} \sum_{k=2}^{N}\sum_{j=1}^{k-1} \delta(x_{k}-x_{j}) 
,
\label{Eq:ArtificialHamiltonian}
\end{equation}
and the correctly normalized ground state $\psi_{0}$ can be put together from \refeq{Eq:FreeBoundState} multiplied by \refeq{Eq:CoMGround}, i.e., $\psi_{0}\equiv\psi_{\textrm{C}}\psi_{\textrm{G}}$, with the sum of the corresponding eigenvalues determining the overall energy $E_{0}$.  Hence, in terms of single particle coordinates
\begin{equation}
\psi_{0}(\vec{x})
=
\left(
\frac{N\gamma}{\pi}
\right)^{1/4}
\exp
\left(
-\frac{\gamma}{2N}
\left[
\sum_{k=1}^{N}x_{k}
\right]^{2}
\right)\psi_{\textrm{G}}(\vec{x}),
\label{Eq:EnvelopeAnsatz}  
 \end{equation}
and, from \refeq{Eq:FreeFullEnergy} plus $\gamma/2$ (the harmonic oscillator zero-point energy),
\begin{equation}
 E_{0} = -\frac{N(N+1)}{24(N-1)} + \frac{\gamma}{2}.
\label{Eq:freeEplusenv}
\end{equation}

This interaction dominated limit does not correspond to any physical system but is a useful starting point for perturbation theory. 

\subsection{Perturbation results}

\begin{figure}
\begin{center}
\includegraphics[width=\linewidth]{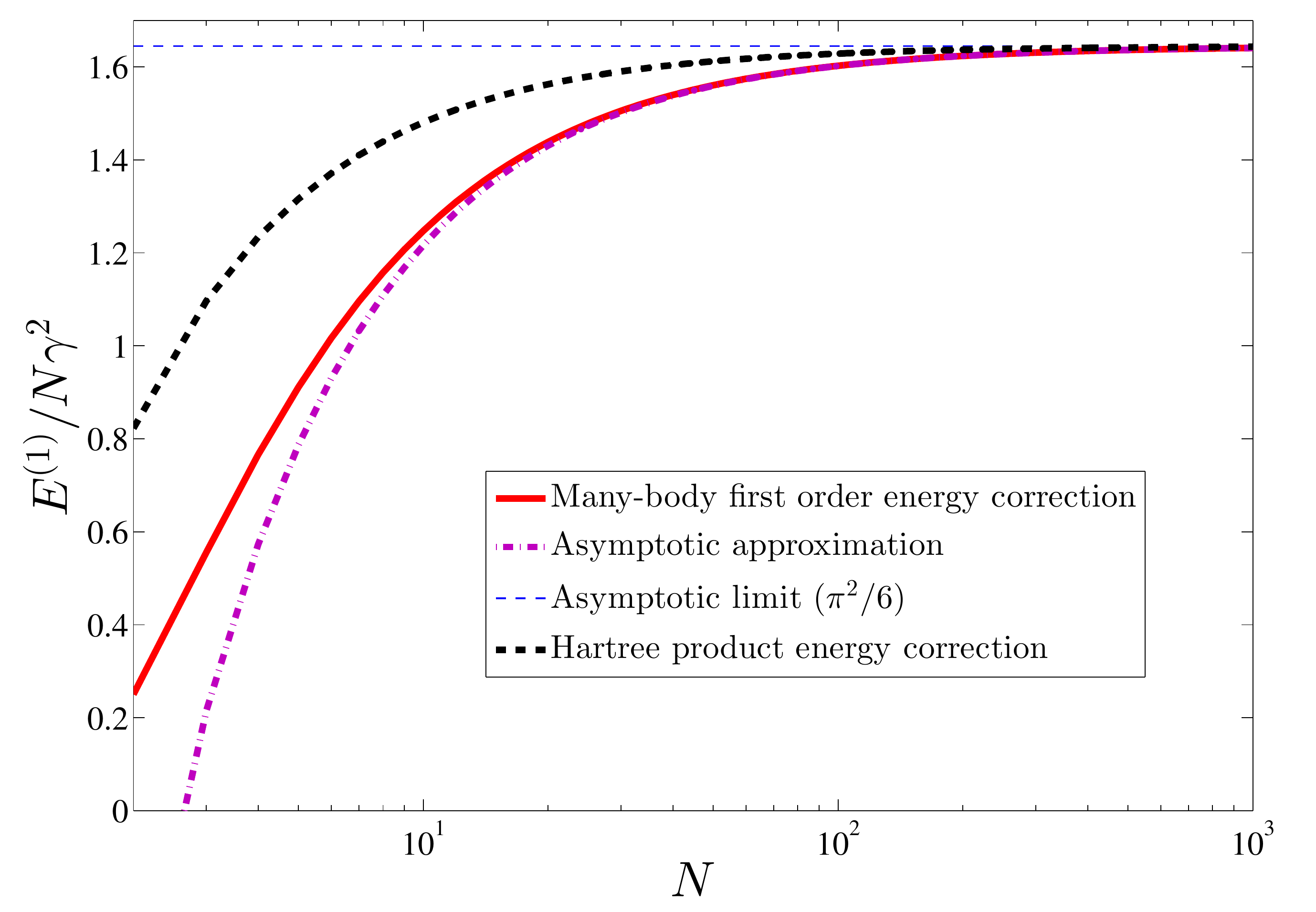}
\caption{(Color online): First-order energy correction per atom for a many-body soliton with the center-of-mass in the lowest eigenstate of a harmonic oscillator given in \refeq{Eq:EnvelopeAnsatz}.  The exact solution given in \refeq{foecor} and its expansion up to next-to-leading order, given in \refeq{eq:foecor2}, begin to agree for $N\gtrapprox 10$ with the latter always underestimating the true value.  Both curves approach the approximate result predicted by the Hartree approximation [\refeq{eq:mfcor}].  The relative difference between different predictions lies below $1 \%$ for $N \gtrapprox 165$.}
\label{fig:energyorders}
\end{center}
\end{figure}

To proceed from the approximated Hamiltonian~(\ref{Eq:EnvelopeAnsatz}), we include the effect of the harmonic trap on the relative degrees of freedom via Rayleigh-Schr\"odinger perturbation theory. The full Hamiltonian~(\ref{Eq:FirstQuantizedH}) can be written as ${H}(\vec{x})={H}_0(\vec{x})+\Delta {H}(\vec{x})$, with
\begin{equation}
\Delta{H}(\vec{x}) =  \frac{\gamma^2}{2} \left[ \sum_{k=1}^N  x_{k}^2  - \frac{1}{N}\left(\sum_{k=1}^N x_{k}\right)^2 \right] \; .
\end{equation}
As $\Delta{H}(x)\propto \gamma^2$, we expect perturbation theory to yield particularly good results in the limit of small $\gamma$. For the first-order energy correction to the ground state,
\begin{align}
 E^{(1)} &= \bra{\psi_{0}} \Delta\hat{H} \ket{\psi_0}  \nonumber \\
	 &=  \int d \vec{x} \; \psi_0(\vec{x})^{*} \Delta{H}(\vec{x})  \psi_0(\vec{x}) \; ,
\end{align}
which also serves as a definition of the bra-ket notation, we find (Appendix \ref{E1calc}): 
\begin{equation}
 E^{(1)} = \gamma^2  \frac{(N-1)^2}{N} \sum_{k =1}^{N-1} \frac{1}{k^2} \; .
\label{foecor}
\end{equation}
The sum in Eq.~(\ref{foecor}) is simply the second Harmonic number, for which the asymptotic behavior in the $N\gg 1$ limit is given by~\cite{Mathematica7010}
\begin{equation}
  \sum_{\ell =1}^{N-1} \frac{1}{k^2} \sim \frac{\pi^2}{6} - \frac{1}{N-1}  + {\cal O}\left([N-1]^{-2}\right) \; .
\end{equation}
 Thus, asymptotically the energy correction goes as
\begin{equation}
 E^{(1)} \sim \gamma^2 \left[\frac{\pi^2}6N -\frac{\pi^2}{3} - 1 + {\cal O} \left(N^{-1}\right)\right]\; .
\label{eq:foecor2}
\end{equation}
For large $N$, this coincides with the result obtained using the free space Hartree solution, given in \refeq{Eq:HartreeSolution}, as an approximation for the ground state (Appendix~\ref{comke})
\begin{align}
 E^{(1)}_{\rm H} &=(N-1) \int_{-\infty}^{\infty} d x \frac{\sech(x/2)^2}{4} \frac{\gamma^2 x^2}{2} 
\nonumber \\ 
&= \gamma^2 (N-1)   \frac{\pi^2}{6}\; .
\label{eq:mfcor}
\end{align}
  These results are displayed in \reffig{fig:energyorders}; for small $N$ there is a large difference between the result predicted by the Hartree product state [Eq.~(\ref{foecor})] and the result predicted by the exact many-body ground state [Eq.~(\ref{eq:mfcor})] of the approximate Hamiltonian~(\ref{Eq:ArtificialHamiltonian}).  There is also a weak number dependence from the Harmonic series in \refeq{foecor}.  However as $N \gg 1$ both methods give the same energy correction per atom, $\pi^2 \gamma^2/6$. For $N=1000$ the relative difference $( E^{(1)}_{\rm H} - E^{(1)})/E^{(1)} \approx  0.0016 $ is already small.

\subsection{Variational minimization \label{sec:varminim}}

In order to improve the value for the ground state energy beyond the first-order-perturbation-theory result~(\ref{foecor}), we use the (normalized) variational ansatz
\begin{equation}
 \psi_{\rm var}^{(\lambda)}(\vec{x}) \equiv  \psi_{\textrm{C}} (x_{\textrm{C}}) {\cal N} \lambda^{(N-1)/2} \exp\left(-\lambda  \sum_{1\le k<j\le N} \frac{\vert x_{k} - x_{j} \vert}{2 [N-1]} \right) \; ,
\label{eq:varpsifn}
\end{equation}
with $\gamma > 0$ and the constant ${\cal N}$ the same as Eq.~(\ref{Eq:FreeBoundState}) \footnote{With just ${\cal N}$ as a prefactor the equation would be normalized with respect to $\lambda \vec{x}$, scaling this out along with $\gamma \to \gamma/\lambda^2$ to keep the center of mass the same, gives the extra factor of $\lambda^{(N-1)/2}$},
\begin{equation}
{\cal N} =
\sqrt{\frac{(N-1)!}{(N-1)^{N-1}}} \; ,
\end{equation}
which is calculated in appendix \ref{sec:norm}.  
Since the center-of-mass wavefunction is unchanged, we will only have a correction to the relative energies, these are calculated in Appendices~\ref{sec:varchanges} and~\ref{E1calc}. The total energy for this wavefunction is
\begin{align}
 \bra{ \psi_{\rm var}^{(\lambda)}} \hat{H} \ket{ \psi_{\rm var}^{(\lambda)}} =\left(2\lambda - \lambda^2 \right) E_{\rm G}  +\frac{E^{(1)}}{\lambda^{2}} +\frac{\gamma^2}2
\label{eq:varenergy}
\end{align}
for the expectation values of each section of the relative Hamiltonian, with $E_{\rm G}$ being the (negative) ground state energy of the free soliton in soliton units given in Eq.~(\ref{Eq:FreeFullEnergy})  and $E^{(1)}$ the first-order correction given by \refeq{foecor}. In order to calculate the energy minimum, the derivative of Eq.~(\ref{eq:varenergy}) with respect to $\lambda$ has to be zero:
\begin{equation}
 (2-2\lambda)E_{\rm G}  - \frac{2E^{(1)}}{\lambda^3} = 0  \; ,
\end{equation}
which (for $\lambda\ne 0$) is equivalent to a 4th order polynomial in $\lambda$
\begin{align}
  \lambda^4-\lambda^3-\kappa= 0  \; ,
\label{eq:varminpoly}
\end{align}
where the constant $\kappa$ is defined as the ratio of the ground state energy and first-order correction
\begin{align}
\kappa &\equiv-\frac{E^{(1)}}{E_{\rm G}} \nonumber \\
&=   \gamma^2\frac{24 (N-1)^3}{(N+1)N^2}\sum_{j=1}^{N-1} \frac{1}{j^2}\;.
\label{eq:kdef}
\end{align}
For fixed $N$, $\kappa\propto \gamma^2$, the value of this prefactor is an increasing function of $N$ with a minimum of $\kappa = 2\gamma^2$ at $N=2$ with an asymptotic limit of [cf.\ Eq.~(\ref{eq:foecor2})]:
\begin{equation}
 \lim_{N \to \infty} \kappa = 4 \pi^2 \gamma^2 \; .
\label{eq:kasNtoinf}
\end{equation}
Thus, $\kappa$ is small for $\gamma\ll 1$. 

Equation~(\ref{eq:varminpoly}) has four roots, only one of which is real and positive, which is the root of interest.  
The exact analytic solution is given in Appendix \ref{sec:quartic}.

If $\kappa \ll 1$, this solution is approximately 
\begin{equation}
\label{eq:lambdaapprox}
\lambda_0 \simeq 1 + \kappa
\end{equation}
 which leads to the minimum in the energy of
\begin{equation}
 E \simeq E_{\rm G} +\frac{\gamma^2}2 + E^{(1)} + \frac{(E^{(1)})^2}{E_{\rm G}} +{\cal O}([E^{(1)}]^3/\vert E_{\rm G}\vert ^2)\; .
\label{eq:smallgamvar}
\end{equation}

As the variational Ansatz~(\ref{eq:varpsifn}) does not affect the center-of-mass part of the wavefunction, calculating the overlap between this variational Ansatz and the state \refeq{Eq:EnvelopeAnsatz} (i.e. the $\lambda = 1$ state) is an interesting physical quantity: its modulus squared is the fraction of the relative wavefunction which is projected to the relative ground state if the trapping potential was turned off quasi-instantaneously (cf.~\cite{Castin2009}). The overlap is given by (see Appendix \ref{sec:varchanges}):
\begin{align}
 \bket{\psi_{\rm var}^{\left(\lambda_0\right)}}{\psi_{0}} &=  \left(\frac{2}{\lambda_0^{1/2} + \lambda_0^{-1/2}}\right)^{N-1} \; .
\label{eq:overlapvar}
\end{align}

Using the approximation~(\ref{eq:lambdaapprox}), the overlap~(\ref{eq:overlapvar}) approximately is $[1+\kappa^2/8 + {\cal O}(\kappa^3)]^{(1-N)}$ and we thus expect the overlap to vanish in the limit $N\to \infty$ for $\kappa> 0$. Rather than investigating the total wavefunction overlap~(\ref{eq:overlapvar}), the $N$th root of Eq.~(\ref{eq:overlapvar}), an effective single particle overlap, is a more suitable value in the limit $N\gg 1$ as it tends to a constant as $N \to \infty$ and is related to comparing two GPE orbitals. Note that for two Hartree-product wavefunctions, the effective single particle overlap would be independent of $N$, but the $N$th root of Eq.~(\ref{eq:overlapvar}) still is $N$-dependent due to the $N$-dependence of $\lambda_0$ [Eq.~(\ref{eq:lambdaapprox}), cf.\ Appendix \ref{sec:quartic}].

\begin{figure}
\begin{center}
\includegraphics[width=\linewidth]{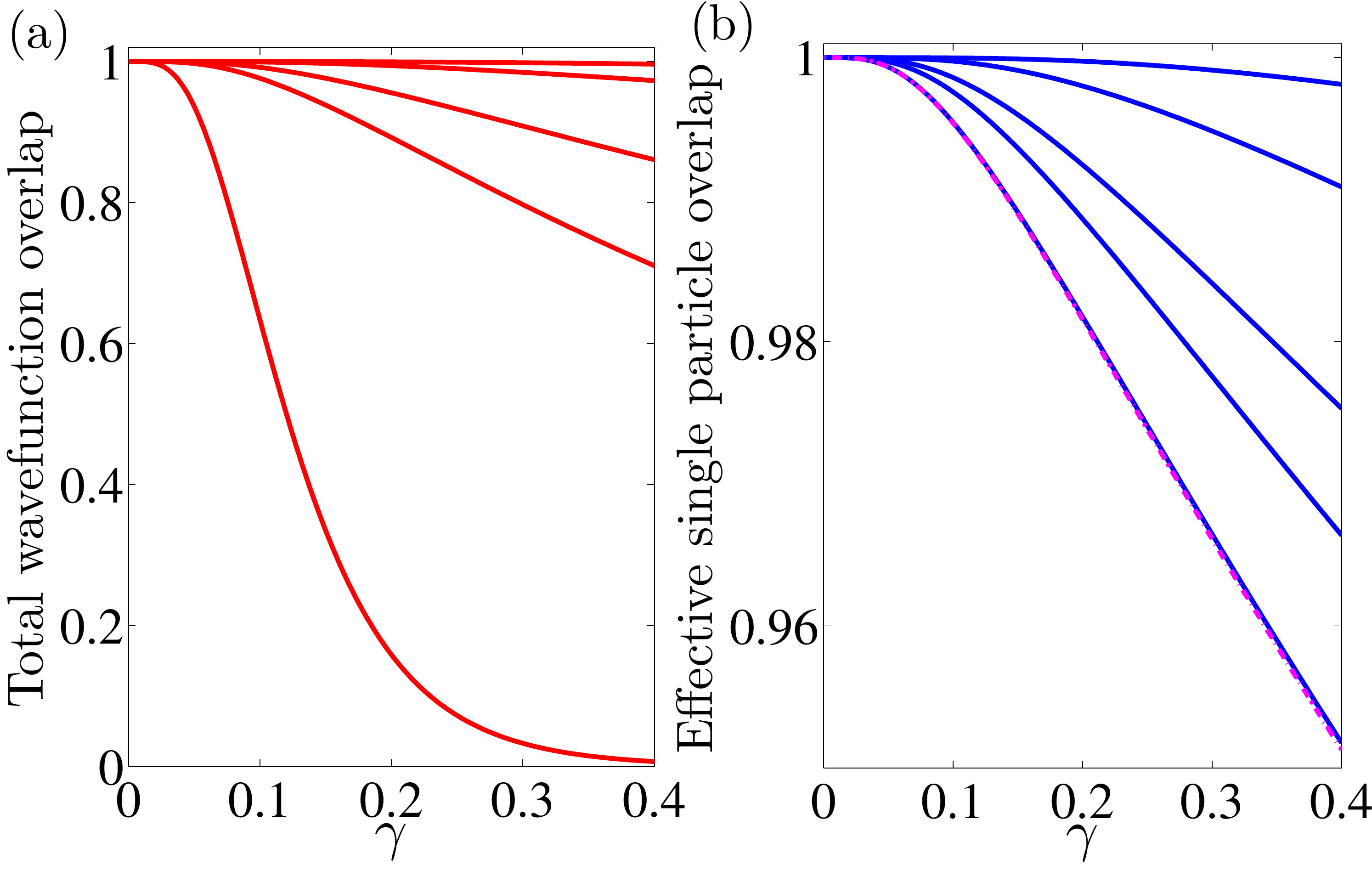}
\caption{(Color online): (a) Total wavefunction overlap, given by \refeq{eq:overlapvar}, (b) effective single-particle overlap, given by the $N$th root of \refeq{eq:overlapvar}, of the variationally obtained solutions for different rescaled trap frequencies $\gamma$ with the free space ground state solution $(\gamma=0)$ with a Gaussian envelope for the center of mass.  Effective single-particle overlap is treated as the $N$th root of the total overlap as for two different product states this is independent of number and equal to the overlap between the single particle wavefunctions.  Bottom to top the solid lines on both graphs correspond to $N=100,10,6,3,2$, the dashed line corresponds to the $N \to \infty$ limit of the variational many-body solution [using $\kappa$ from \refeq{eq:kasNtoinf}] and is very close to the $N=100$ line.} 
\label{overlapvar}
\end{center}
\end{figure}

Figure \ref{overlapvar} (a) shows the overlap~(\ref{eq:overlapvar}) as a function of $\gamma$ for various particle numbers. For $\lambda_0$, the exact value given in Appendix~\ref{sec:quartic} was used.    
As expected, the $N$-dependence is quite strong.
Figure \ref{overlapvar}~(b) shows the $N$th root of the overlap~(\ref{eq:overlapvar}), i.e.\ the effective single-particle overlap. The effective single-particle overlap is larger than $0.99$ for $\gamma \lesssim 0.15$ for all $N$, indicating that $\psi_0 (\vec{x})$ from ~(\ref{Eq:EnvelopeAnsatz}) is still a good description in this parameter regime and the trap has had little effect on the internal degrees of freedom. The limit $N\to\infty$ is nearly reached for particle numbers as low as $N=100$ [note that in panel (a), the limit $N\to\infty$ would lie on the coordinate axes].

\newpage

\vspace*{1cm}

\section{Computational methods including a harmonic potential \label{sec:hop}}
\subsection{Overview}
While the focus of the previous section lies on the case of small $\gamma$, the numerical methods introduced in this section  work well for $\gamma \gtrsim 0.16$.

\subsection{Computation procedure}  

We expand the field operator over the set of Hermite functions 
\begin{equation}
\varphi_k(Wx) = \sqrt{\frac{W}{k!2^k\pi^{1/2}}} H_k(W x) \exp\left(-W^2 x^2/2\right) \; , 
\label{eq:Wbasis}
\end{equation}
where $H_k$ are the Hermite polynomials, giving $\hat{\Psi}(x) = \sum_k \varphi_k(x) \hat{a}_k$.  The Hamiltonian~(\ref{Eq:ScaledSecondQuantizedH}) in this basis can be split into three separate parts:
\begin{align}
 \hat{H}_{\rm K} =  &\frac{W^2 }{4} \sum_k \left[ (2k+1) \hat{a}^{\dagger}_k\hat{a}_k \right. \nonumber \\
   &-  \left. \sqrt{(k+1)(k+2)}(\hat{a}^{\dagger}_{k+2}\hat{a}_k + \hat{a}^{\dagger}_k\hat{a}_{k+2})\right] \ \; ,
\label{eq:HK}
\end{align}
the kinetic Hamiltonian,
\begin{align}
 \hat{H}_{\rm P} =  &\frac{\gamma^2 }{4W^2} \sum_k (2k+1) \left[ \hat{a}^{\dagger}_k\hat{a}_k  \right. \nonumber \\
\phantom{\sum_k (2k+1)}  &+ \left. \sqrt{(k+1)(k+2)}(\hat{a}^{\dagger}_{k+2}\hat{a}_k + \hat{a}^{\dagger}_k\hat{a}_{k+2})\right] \ \; ,
\label{eq:HP}
\end{align}
the potential Hamiltonian, and
\begin{equation}
 \hat{H}_{\rm I} = - \frac{W}{N-1} \sum_{k \ell mn} f_{k \ell mn} \hat{a}^{\dagger}_k\hat{a}^{\dagger}_{\ell}\hat{a}_m \hat{a}_n \; .
\label{eq:HI}
\end{equation}
the interaction Hamiltonian.  The factor of $f_{k \ell mn}$ is the integral of four Hermite functions (with $W$ set to unity) over all space, i.e.\
\begin{equation}
f_{k \ell mn} =  \int dx \; \varphi_k(x) \varphi_{\ell}(x) \varphi_m(x) \varphi_n(x)  \; .
\label{fklmn1}
\end{equation}
\begin{widetext}
Here, the functions are also real, so there is no need to take the complex conjugates.  This can be calculated exactly in terms of gamma functions $\Gamma(x)$ and a standard hypergeometric function $_3F_2$ evaluated at unity \cite{Busbridge1948}
\begin{align}
f_{k \ell mn} =  & \frac1{\sqrt{2}\pi^2} \sqrt{\frac{m!}{k!\ell!n!(m-n)!}}\Gamma([k+\ell-m+n+1]/2)\Gamma([k-\ell+m-n+1]/2) \nonumber \Gamma([-k+\ell+m-n+1]/2) \nonumber \\
       &\times \;_3 F_2([-n,(m-n+k-\ell+1)/2,
 (m-n-k+\ell+1)/2];
[1+m-n,(m-n-k-\ell+1)/2],1) \; .
\end{align}
\end{widetext}

Without interactions, the ideal gas Hamiltonian is given by
\begin{equation}
\hat{H}_{\rm ideal} =  \hat{H}_{\rm K} +  \hat{H}_{\rm P}.
\label{eq:Hideal}
\end{equation}
For  $W=\sqrt{\gamma}$, the basis states~(\ref{Eq:EnvelopeAnsatz}) are eigenstates of the non-interacting Hamiltonian.
 The total Hamiltonian can therefore be expressed as
\begin{equation}
 \hat{H} =  \gamma \sum_k \left(k+\frac{1}{2}\right) \hat{a}^{\dagger}_k\hat{a}_k  - \frac{\sqrt{\gamma}}{2(N-1)} \sum_{k \ell mn} f_{k \ell mn} \hat{a}^{\dagger}_k\hat{a}^{\dagger}_{\ell}\hat{a}_m \hat{a}_n \; ,
\label{eq:compham}
\end{equation}
and we refer the ground state of this as $\ket{\psi_{\rm g}(\gamma)}$ and the ground state energy as
\begin{equation}
\bra{\psi_{\rm g}(\gamma)} \hat{H} \ket{\psi_{\rm g}(\gamma)} = E_{g} (\gamma) \; .
\label{eq:trapgs}
\end{equation}

\subsection{Truncation and projection to center-of-mass excitation basis}

In order to do computations we must only use a finite basis set, which will introduce the inaccuracy. This is discussed in Appendix \ref{truncate}. Essentially all possible states for which the eigenenergy related to the Hamiltonian~(\ref{eq:Hideal}) lies below an energy cut-off $E_{\rm cut}$ are included, we refer to this set as the `truncated basis'. 

The brute force procedure would now be to calculate all the matrix elements of the interaction Hamiltonian (\ref{eq:HI}) using this truncated basis and add the matrix of energies of the kinetic and potential Hamiltonians and diagonalize this to get the eigenstates and energies.  However we can reduce the size of the truncated basis set used in this computation by recalling from \refeq{Eq:HamiltonianC} that the center-of-mass Hamiltonian commutes with the relative Hamiltonian and thus they have separate eigenstates. 

Inspired by the ladder operator treatment of a single particle in a harmonic oscillator, we define
\begin{align}
A^{\pm}(x_{\rm C}) = \sqrt{\frac{1}{2N\gamma}} \left(N \gamma x_{\rm C}  \mp \pd{x_{\rm C}} \right) \; .
\end{align} 
Noting we can express the center of mass Hamiltonian as $H_{\rm C}(x_{\rm C}) = \gamma(A^{+}(x_{\rm C}) A^{-}(x_{\rm C}) + 1/2)$, as is the case of the single particle ladder functions.  Moving to second quantization, we can equivalent operators in terms of creation and annihilation operators in our basis (for $W = \sqrt{\gamma}$) via~\cite{HaugsetHaugerud1998} 
\begin{equation}
\hat{A}^{-} = \sum_k \sqrt{k+1}\hat{a}^{\dagger}_{k}\hat{a}^{\;}_{k+1}  \;.
\label{eq:modecreation}
\end{equation}
where $\hat{A}^{+} = (\hat{A}^{-})^{\dagger}$.  These satisfy $[\hat{H}_0,\hat{A}^{\pm}]=\pm \gamma \hat{A}^{\pm}$ and thus energy levels spaced in units of $\gamma$; they also commute with the interaction Hamiltonian $[\hat{H}_I,\hat{A}^{\pm}]=0$ as required.  Note that for $N=1$, $\hat{A}^{+}$ acting on the ground state can be used to construct all the eigenstates of the system. 
We then can express the center-of-mass Hamiltonian as 
\begin{align}
\hat{H}_{\rm C} &= \frac{\gamma}{N}  \hat{A}^{+} \hat{A}^{-} + \frac{\gamma}{2} \; ,
\label{eq:comham}
\end{align}
In order to use this property to reduce the basis, we must first transform our truncated basis ($E \le E_{\rm cut}$) to basis states which are eigenstates of the center-of-mass Hamiltonian, this procedure is detailed in Appendix \ref{sec:com} along with the reduction in basis size it achieves. 

We keep only eigenstates of $\hat{H}_{\rm C}$ with eigenvalue $\gamma/2$ (center-of-mass ground state) and project the Hamiltonian (\ref{eq:compham}) from the truncated basis to this new and reduced basis. 
For high $N \ge E_{\rm cut}/\gamma-N/2$, the reduction asymptotes to $\pi/\sqrt{6(E_{\rm cut}/\gamma-N/2)}$ for  $E_{\rm cut}/\gamma - N/2\gg 1$. For details see Appendix \ref{sec:basisred}.

\subsection{Using different-width Hermite functions}

Using functions with a $W =\sqrt{\gamma}$, such that they are eigenstates of $\hat{H}_{\rm ideal}$, is not desirable in the $\gamma \to 0$ limit because the basis will consist of states much wider than the wavefunction we are using them to construct. 
For an infinite basis, the ground state should be independent of the basis used to describe the system (in our case, it should be independent of the value of $W$). For numerical calculations, the basis will be finite and thus some choices of $W$ are better than others. In Sec.~\ref{sec:results}, we will calculate the ground state for $\gamma=0$ in order to determine the optimal value for $W$ to be used in the calculations.

For arbitrary $W$, the Hamiltonian now reads:
\begin{align} 
 \hat{H} =&   \sum_{k} \left[\frac{W^2 + \gamma^2 W^{-2}}{2} \left(k+\frac{1}{2} \right) \hat{a}^{\dagger}_k\hat{a}_k \right.  \nonumber \\
&+ \left. \frac{\gamma^2  W^{-2}-W^2}{4} \sqrt{(k+1)(k+2)}(\hat{a}^{\dagger}_{k+2}\hat{a}_k + \hat{a}^{\dagger}_k\hat{a}_{k+2})\right]\nonumber \\
&-\frac{W}{2 (N-1) } \sum_{k \ell mn} f_{k \ell mn} \hat{a}^{\dagger}_k\hat{a}^{\dagger}_l\hat{a}_m \hat{a}_n \; ,
\label{eq:compham2}
\end{align} 
which includes extra mixing terms in the ideal gas Hamiltonian~(\ref{eq:Hideal}).  This causes a fairly significant issue in that it is no longer possible to exactly separate center-of-mass eigenstates in this basis, meaning the full basis would need to be used in order to achieve the center-of-mass ground state. 
Using this method with just a truncated Hilbert space and no projection to the sub space with zero center-of-mass excitation 
would make achieving convergence painfully slow.  The solution to this is therefore to reduce the basis in the same way as before, but accept that the center-of-mass wavefunction we end up with is given by
\begin{align} 
f_{\rm C}(x_{\rm C}) = \sqrt{\frac{W}{\pi^{1/2}}} \exp\left(-N\frac{W^2 x_{\rm C}^2}{2} \right) \; ,
\end{align} 
which is not an eigenstate and has energy \mbox{$E_{\rm C} = (W^2 + \gamma^2 W^{-2})/4$} rather than the true $\gamma/2$, thus we know the true ground state is the wavefunction we obtained, multiplied by $\sqrt{\sqrt{\gamma}/W}\exp\left([\gamma-W^2] N x_{\rm C}^2/2\right)$.
This approach has the huge advantage that, if $W$ is kept constant, the occupation of the basis states for the ground state should change very little as $\gamma \to 0$ where they will tend to the solutions on the infinite line.   

\subsection{Numerical ground states within the GPE approximation \label{sec:GSHartree}}

Within the GPE approximation, we can obtain the ground state by solving \refeq{Eq:ScaledGPE} as the ground state is the only stationary state of the system. 
The method used here is to again expand over a finite basis set of Hermite functions of arbitrary width scaling $W$
\begin{equation}
  \phi(x) = \sum_{k=0}^{\eta} c_k \sqrt{W} \varphi_k (Wx) \; ,
\end{equation}
then to produce a set of $\eta+1$ nonlinear equations in the coefficient set $\mathbf{c}$ (which will be real), by integrating \refeq{Eq:ScaledGPE} multiplied by $\varphi_k(x)$ over all space, for $k=\{0,\ldots,\eta\}$, giving
\begin{align}
 0 =  &-\mu c_k + \frac{W^2 + \gamma^2 W^{-2}}{2} (k+1/2) c_k \nonumber \\
&+ \frac{\gamma^2  W^{-2}-W^2}{4} \left( \sqrt{(k+1)(k+2)}c_{k+2} + \sqrt{k(k-1)}c_{k-2} \right) \nonumber \\
&-\frac{W}{2 (N-1) } \sum_{\ell, m,n=0}^{\eta} f_{k \ell mn} c_{\ell} c_{m} c_{n} c_{k} \; .
\end{align}
There is also an $(\eta+2)$th equation, relating to the normalization $\sum_k \vert c_k \vert^2 = 1$.  Denoting the vector with an equation at each position as $\mathbf{F(\mathbf{c})} $, we wish to solve $\mathbf{F} = \mathbf{0}$. We use Newton's method (as in~\cite{HaugsetHaugerud1998}) to iteratively solve for $\mathbf{c}$, via
\begin{equation}
 J(\mathbf{c}^{(n)}) \; (\mathbf{c}^{(n+1)}-\mathbf{c}^{(n)})  =  \mathbf{F}({\mathbf c}^{(n)}) \; ,
\end{equation}
where $J$ is the $\eta+1$ by $\eta+2$ Jacobian matrix associated with $F$.  $\eta$ is increased until convergence is achieved.

\section{Effects of harmonic confinement\label{sec:results}}

\subsection{Ground state energy}
Using the methods from the previous two sections, we investigate the effect an external potential has on the relative component of the ground state $\ket{\psi_{\rm g}}(\gamma)$ [cf.~\refeq{eq:trapgs}].  This is important to quantify how soliton-like the state is, along with what excitations can be expected if the state is released quasi-instantaneously from the potential.
  Such dynamics have already been considered using the GPE in~\cite{Castin2009}. 

Figure \ref{fig:energy} shows $\Delta E /N$, the energy difference per atom between the numerically calculated ground state energy $E_{\rm g}(\gamma)$ and the ground state energy of the artificial Hamiltonian (\ref{Eq:EnvelopeAnsatz}) given by \refeq{Eq:freeEplusenv}, for a range of $\gamma$ and $N$ values.  It is produced by calculating the ground state energy via the three numerical methods, namely exact diagonalisation in a basis of hermite functions with either optimized widths for weak trapping (shown in table \ref{table}), or widths which are eigenstates of the non interacting problem, and variational minimization, for a range of $\gamma$ and taking the smallest value.  This is because, due to the variational principle, all of these techniques produce only values greater than or equal to the ground state energy, hence the lowest is the best estimate.  

\subsection{Universal behaviour}

\begin{figure*}
\begin{center}
\includegraphics[width=0.95\linewidth]{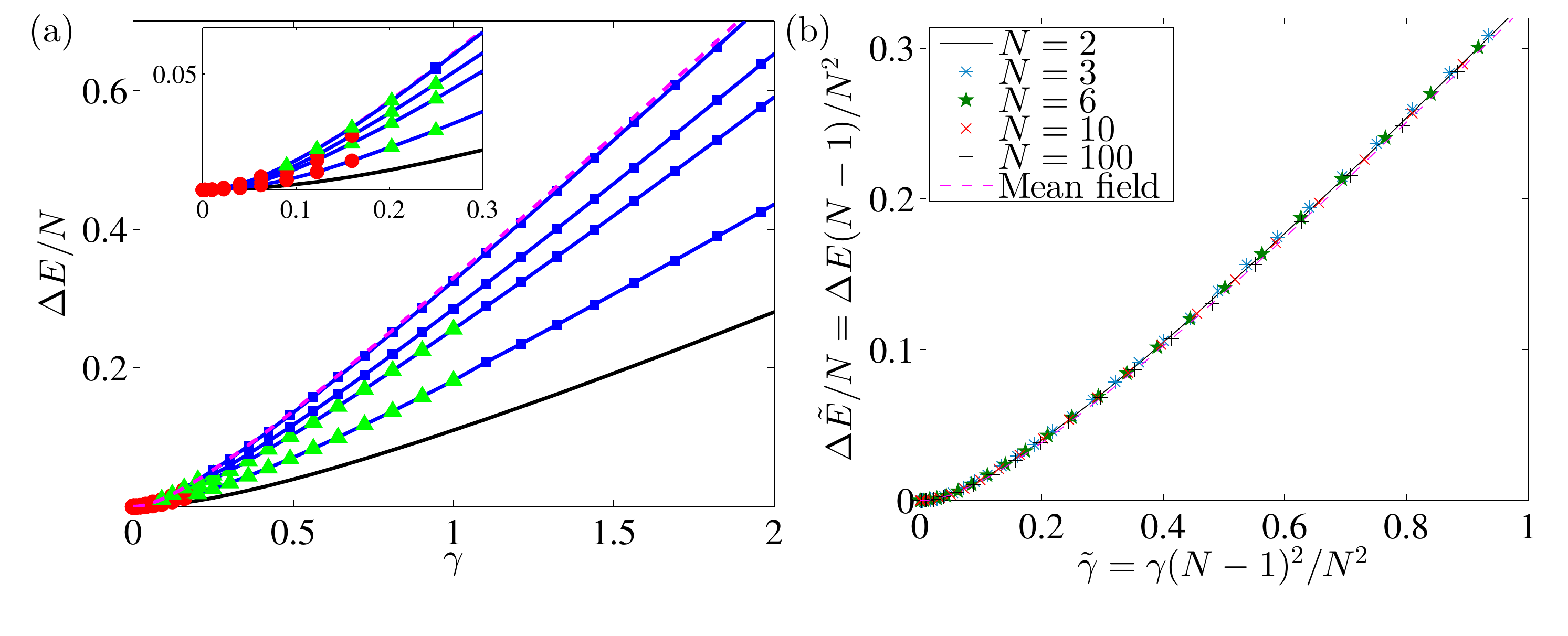}

\caption{(Color online): Difference between the ground state energy per atom and the energy of a free many body ground state with a Gaussian center-of-mass profile, (a) is in terms of $E_0/N$ [as defined in \refeq{Eq:freeEplusenv}] as a function of rescaled trapping strength $\gamma$ and (b) with a rescaled $\tilde{\gamma} = \gamma (N-1)^2/N^2$ and $\Delta \tilde{E} = \Delta E (N-1)/N$ .  From bottom to top the lines on (a) are $N=2,3,6,10,100$ and the dotted top line is the GPE prediction (which will agree with the many body results as $N\to \infty$) outlined in Sec.~\ref{sec:GSHartree}. The markers indicate a point on the line generated by different methods, (red) circles use the variational solution~\refeq{eq:varenergy}, (green) triangles use the fixed width basis sets to find the lowest eigenvalue of  the Hamiltonian (\ref{eq:compham2}), cf.\ table~\ref{table}. (Blue) squares are obtained using the basis of eigenstates of the non interacting Hamiltonian (\ref{eq:Hideal}) to find the lowest eigenvalue of \refeq{eq:compham}.  The $N=2$ line is plotted using the exact solution [\refeq{Eq:TwoBodyEnergy}] detailed in Sec. \ref{sec:2 bosons} and the mean field line is obtained by the method explain in Sec. \ref{sec:GSHartree}.  The inset shows a zoom of the low $\gamma$ section, demonstrating the initial quadratic dependence on $\gamma$.  Figure (b) shows the universal behaviour present using rescaled units, this is the same data as in (a), however the numerical lowest eigenvalues are plotted as points to make them visible.}

\label{fig:energy}
\end{center}
\end{figure*}

If we consider instead a rescaling $\tilde{\gamma} = \gamma (N-1)^2/N^2$ and $\Delta \tilde{E} = \Delta E (N-1)/N$ (i.e. converting to a unit system in which $g_{1 \rm D}N = 1$ as opposed to $g_{1 \rm D} (N-1) = 1$), a more universal behaviour is present in $\Delta \tilde{E}$, with little number dependence as shown in fig.~\ref{fig:energy} (b).  To see this analytically, we note that for $\tilde{\gamma} \ll 1$, our variational result of~\refeq{eq:smallgamvar} for the energy is applicable. $\Delta E$ is obtained by subtracting the factor of $E_G +\gamma^2/2$, then converting to our rescaled units we have
\begin{align} 
 \frac{\Delta{\tilde{E}}}{N} \approx \; &\tilde{\gamma}^2 \frac{ N }{N-1} \sum_{k=1}^{N-1} \frac{1}{k^2} \nonumber \\
 &- \tilde{\gamma}^4 \frac{ 24 N^3 }{(N^2-1)(N-1)}\left[\sum_{k=1}^{N-1} \frac{1}{k^2}\right]^2 + { \cal O}(\tilde{\gamma}^6)\; .
\end{align} 
the $N$ dependent factor of order $\tilde{\gamma}^2$ (which is the rescaled first order energy correction) is $2$ for $N=2$ and decreases monotonically to $\pi^2/6 \approx 1.6$ as $N \to \infty$, hence for very small $\tilde{\gamma}$, the $N=2$ line is largest, but the difference is very small.  
The  order $\tilde{\gamma}^4$ term has negligible number dependence and so is unlikely effect the ordering of these lines within for the range of variational models validity.  On the other end of the scale, as $\gamma \to \infty$ (the trap dominated system) we can neglect interactions in \refeq{Eq:ScaledSecondQuantizedH}, giving a ground state of a product of Gaussians of width $1/\gamma$, subtracting the center-of-mass energy gives $\Delta E \to \gamma(N-1)/2N + {\cal O}(\sqrt{\gamma})$ or, in our rescaled units, $\Delta \tilde{E} \to \tilde{\gamma}/2 + {\cal O}(\sqrt{\tilde{\gamma}})$ and so to leading order, the $N$ dependence vanishes.

\begin{table}
    \begin{tabular}{ | p{1cm} | p{1cm}  | p{1cm}  | l | } 
    \hline 
     $\; N$   & $\; \eta$   & $\;  W$   &$\;$ Reduced basis size $\;$ \\ \hline \hline
    $\; 3$ & $\; 84$ & $\; 2$ & $\; 631$ \\ \hline
    $\; 6$ & $\; 38$ & $\; 1$ & $\; 3009$\\ \hline
    $\; 10$ & $\; 28$ & $\; 0.5$& $\; 2534$\\  \hline
    $\; 100$ & $\; 24$ & $\; 0.5$& $\; 1575$ \\  \hline
    \end{tabular}
\caption{\label{table}This table shows the parameters used in calculating the graph in Fig.~\ref{fig:energy}, where $N$ is atom number, $\eta$ is the cut-off and $W$ is the width taken for the fixed width calculations (coarsely chosen to minimize the ground-state energy at $\gamma$ = 0).  The reduced basis size is the number of states (with zero center-of-mass excitation) used in the exact diagonalisation of the Hamiltonian.  The basis are chosen to be a reasonable computational size, however the numerics are less reliable for small $\gamma$}
\end{table}

\subsection{The classical soliton limit \label{sec:agreement}}  

As shown in Fig.~\ref{fig:energy}, for low $\gamma$ the variational ansatz~(\ref{eq:varpsifn}) gives the best estimate for the ground state energy of all the methods used in this paper. For low enough $\gamma$, this variational ansatz in turn is very close to the product of the free many-particle solution with a Gaussian center-of-mass wavefunction~(\ref{Eq:EnvelopeAnsatz}) (see Fig.~\ref{overlapvar}). In the limit of small $\gamma$, the integral
\begin{equation}
 {\cal B} = \int_{-\infty}^{\infty} dx_1\ldots\int_{-\infty}^{\infty} dx_N \; \psi_{0}(\vec{x}) \psi_{\rm H}(\vec{x})  \; 
\label{eq:defB}
\end{equation}
can thus be used to investigate deviations of the Hartree-product wavefunction~(\ref{Eq:HartreeSolution}) from the true many-particle ground state (which is well approximated by $\psi_0(\vec{x})$ for $\gamma \ll 1$). As was the case for Fig.~\ref{overlapvar}, the effective single particle overlap ${\cal B}^{1/N}$ will also be considered as we are interested to see how well the wavefunction is described by a product state, and when comparing two product states with different single particle wavefunctions, this quantity is constant with changes to number.   

\begin{figure*}

\includegraphics[width=0.9\linewidth]{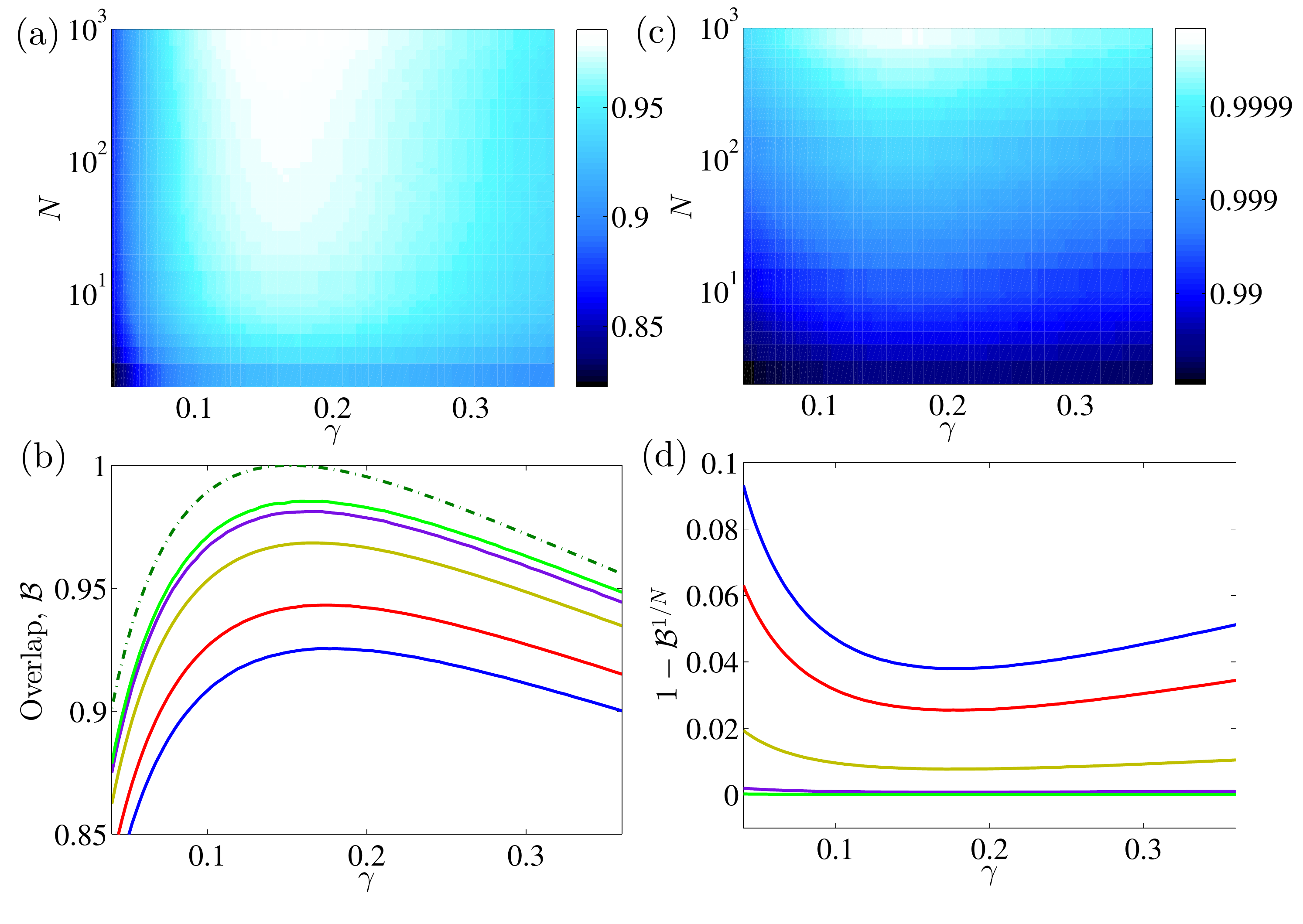}

\caption{(Color online): (a) 2D projection of overlap, ${\cal B}$ as defined in \refeq{eq:defB}, between many-body free state with a Gaussian envelope (of width $\propto 1/\gamma$) given in \refeq{Eq:EnvelopeAnsatz} and the mean field soliton solution, given in \refeq{Eq:HartreeSolution}, for a range of $N$ and $\gamma$. (b) shows Horizontal slices through (a), and the dash-dotted line is the analytic estimate, based purely on center-of-mass position uncertainty, given by \refeq{eq:gaussgauss}. (c) shows effective single particle overlap ${\cal B}^{1/N}$ and (d) shows the residuals $1-{\cal B}^{1/N}$ for given $N$ values again via slices through (c).   
The solid lines in the lower figure (b) correspond to $N=2,3,10,100,1000$ in that order from bottom to top, and this ordering is reversed for figure (d).  
As expected, the effective single particle overlap plot (c) show a rapid convergence to unity as $N$ increases. [(a) and (b)] suggest that most of the $\gamma$ dependence in ${\cal B}$ is due to the effective ``center-of-mass width'' of the Hartree product solution, since the shape of each overlap curve is similar, besides a small offset, to the dash-dotted line.  This indicates that a Hartree soliton is a very good approximation to the many body solution if the center-of-mass wavefunction is also localized.}
\label{fig:mfvsmb}

\end{figure*}  

In order to make an educated guess about what range of $\gamma$ will give a large overlap, we look at the expectation value of the square of the center-of-mass location over the Hartree-product wavefunction (Appendix~\ref{comke}):
\begin{equation}
 \bra{\psi_{\rm H}} x_{\rm C}^2 \ket{\psi_{\rm H}} = \frac{\pi^2}{3N} \; .
\label{eq:sdhartreecom}
\end{equation}
This value is identical to the variance of the center of mass, as both the many body and Hartree states are centered about $x=0$. A variance calculation can also be performed for \refeq{Eq:EnvelopeAnsatz}, this is particularly simple as the center-of-mass is explicitly separate and is given by
\begin{equation}
 \bra{\psi_{0}} x_{\rm C}^2 \ket{\psi_{0}} = \frac{1}{2 \gamma N} \; .
\label{eq:sdmbcom}
\end{equation}
As the Hartree product state is uncorrelated, for a large enough $N$ the distribution associated with the center-of-mass location will therefore tend to a Gaussian (with variance of $\pi^2/3N$) via the central limit theorem.  We therefore consider an effective center-of-mass wavefunction that is the square root of this distribution, yielding the overlap integral
\begin{align}
I(\gamma) &= \int^{\infty}_{-\infty}\;  dx_{\rm C} \; \left(\frac{3N^2\gamma}{2\pi^{4}}\right)^{1/4}\exp\left[-N\frac{(\gamma+3/2\pi^2) x_{\rm C}^2}{2}\right] \nonumber\\
&= \frac {(24\pi^2\gamma)^{1/4}}{\sqrt {2\,
\gamma\,{\pi }^{2}+3}}
 \; ,
\label{eq:gaussgauss}
\end{align}
which reaches its maximum $I\left(\gamma_{\rm max}\right)=1$ for
\begin{align}
\gamma_{\rm max} &= \frac 3{2\pi^2}\nonumber \\
&\simeq 0.15 \; .
\label{eq:gammamax}
\end{align}
As the above analysis focuses on the center-of-mass part of the wavefunction, and thus \refeq{eq:gaussgauss} is likely to overestimate the overlap ${\cal B}$ as defined in \refeq{eq:defB}, \refeq{eq:gaussgauss} also predicts a $\gamma \to 0$ behaviour of the form $I(\gamma) \sim c \gamma^{1/4}$ with $c$ a constant.  
Figure~\ref{fig:mfvsmb} shows a numerical calculation of ${\cal B}$ for a range of $\gamma$ and $N$, the integration is performed via Monte Carlo methods, i.e.\ weighted sampling using random variables with a $\sech(x/2)^2/4$ distribution (obtained via the ziggurat algorithm~\cite{MarsagliaTsang2000}) until a standard error of $< 10^{-4}$ was obtained.  

It can be seen from \reffig{fig:mfvsmb} (a) and (b), that maximum overlap occurs just slightly above $\gamma = 0.16$ [close to the analytic estimate~(\ref{eq:gammamax})] and improves as $N$ increases.  Based on our previous discussion of effective center-of-mass width, the top value should relate to the overlap of the relative degrees of freedom, although this is not well defined.  Graphs (c) and (d) show the $N$th root of (a) and (b), effectively overlap at the level of single particles, which tends extremely rapidly to unity as $N$ increases for any $\gamma$ over the range shown.  

A useful point that this high overlap implies is that the many-body state \refeq{Eq:EnvelopeAnsatz}, is extremely well approximated by the Hartree product state if the center-of-mass envelope squared is approximately the statistical distribution that would arise from taking the mean of the $N$ independent probability distributions $\vert \phi \vert^2$ [with $\phi$ given in \refeq{Eq:ClassicalSoliton}] associated to single atom positions in the product state.  For this reason the Hartree product state would be expected to well approximate the ground state of the system, even at a many-body level, if the center-of-mass envelope is localized to the size of this distribution (i.e. $\gamma \approx 3/2\pi^2$) by the potential.  The converse to this is also true, an initial condition that is given by~\refeq{Eq:HartreeSolution} is well approximated by~\refeq{Eq:EnvelopeAnsatz} with $\gamma\approx 3/2\pi^2$.  This could be used to estimate center-of-mass position uncertainty of a state, initially given by a Hartree product wavefunction, as it evolves in time, using known results for the spreading of Gaussian wavepackets.   

We also consider how these many-body effects would affect experimental observations. From a measurement the atomic density, one could use the mean of this signal to determine the center-of-mass of the system.  If the state of the system is well approximated by~\refeq{Eq:EnvelopeAnsatz}, the observed location would vary shot to shot with a probability distribution given by $\vert \psi_{\rm cm}(x_{\rm C})\vert^2 \propto \exp(-N\gamma x_{\rm C}^2)$ (combined with any experimental uncertainties associated with density measurement).  For $\gamma < \gamma_{\rm max}$, this distribution would be wider than one would expect using the product approximation, most notably if the center-of-mass wavefunction is wider than a classical soliton width $1/\sqrt{\gamma N} \gtrapprox 1$, this jumping effect would be most clearly visible.  Non zero temperature would further increase this effect, by introducing a statistical mixture of excited states of the center of mass.  As a purely mechanical analogy, one could think of taking a photo of a swinging pendulum at a random time, the shape always looks the same but its position appears random.

\section{Conclusions \label{sec:conc}}



We study a 1D system of identical Bosons with attractive contact interactions, a Lieb-Liniger(-McGuire) gas, in the presence of a harmonic trapping potential.  We present variational and numerical many-body calculations, in both cases making use of the separability of the center-of-mass Hamiltonian to split the problem into relative and center-of-mass degrees of freedom.  We use a unit system such that the Hartree soliton length is set to unity ($\hbar = m = g(N-1)=1$), leaving two parameters, the number of atoms, $N$, and $\sqrt{\gamma}$, the dimensionless ratio between the Hartree soliton and harmonic oscillator lengths.

Our key results are firstly that we have derived a first order energy correction to the ground state of the relative degrees of freedom from the introduction of a harmonic oscillator potential [given in \refeq{foecor}], which is used in a variational minimization technique.  This is proportional to $\gamma^2$ and the correction per atom tends to the mean field prediction from below, the relative difference is less than $1\%$ for $N > 165$. 

Secondly we have determined the validity range of $\gamma$ of our many body ansatz, consisting of the free many-body ground state with a Gaussian envelope as given in \refeq{Eq:EnvelopeAnsatz}.  Essentially as the trapped ground state deviates from this it becomes less ``soliton like'', we quantify this with the ``effective single particle overlap'', given by the $N$th root of the overlap between a variationally obtained ground state and our ansatz.  For $N$ large, this overlap is greater than $0.99$ for $\gamma \lesssim 0.16$.  Numerical calculations of energy in the strongly trapped region, $\gamma>1$, indicate energies are still considerably lower than the non interacting case.

Thirdly we show, via a numerical investigation of overlap between the free Hartree product solution and the free many-body ground state with a Gaussian envelope [given in \refeq{Eq:EnvelopeAnsatz}] describing the center-of-mass wavefunction, that the two wavefunctions can have high agreement, even at a many-body level.  This high overlap occurs when the modulus square of the envelope function matches the probability distribution, associated with the Hartree product, for the center-of-mass position, which occurs when $\gamma \approx 0.16$.  However, current experiments with bright matter-wave solitons are such that the center of mass is localized to much less than a soliton width, indicating this is unlikely to be an observable effect.
 
In addition to these physical results, we outlined a numerical method for computing many body eigenstates, this uses a basis set of harmonic oscillator eigenstates, truncated at a particular energy.  This is then project a into a subspace of states with the center-of-mass wavefunction in a specific state, using the ladder operator for center-of-mass excitation.  This makes use of the separability and achieves a reduction in the size of the basis set required by a factor of $\pi/\sqrt{6E_{\rm cut}}$ (where $E_{\rm cut}$ is a cut-off energy) or better, greatly improving speed of the diagonalization and allowing us to investigate the internal degrees of freedom separately.  
\acknowledgments

We would like to thank T.P. Billam and J. Brand for discussions, and the UK EPSRC for funding (Grant No. EP/G056781/1) and the Jack Dodd Centre (S.A.G.) for support.  

\begin{appendix}

\section{The 1D free system and the full eigenspectrum via the Bethe Ansatz \label{betheans}}

We briefly recapitulate aspects of the treatment of the attractively interacting Lieb-Liniger gas \cite{McGuire1964,LaiHaus1989B} in order to set notation within our chosen system of units.

In order to find the solution to the ground state of \refeq{Eq:ScaledFirstQuantizedH} with $\gamma = 0$, we note that the wavefunction in the region $x_1<x_2<\ldots<x_N$ is solved by 
\begin{equation}
 \psi(\vec{x}) = \frac{1}{\sqrt{N}}\sum_{\{\cal P\}} A({\cal P}) \exp\left(i \sum_{k=1}^N p_{{\cal P}(k)} x_{k}\right),
\end{equation}
where a sum over all permutations of the set ${\cal P} = \{1,\ldots,N\}$ is performed to make it symmetric, the energy eigenvalue is thus simply equal to $E = \sum_k p_k^2/2$. Each permutation has a coefficient associated with it that is linked to the boundary conditions when $x_{k} = x_{k+1}$, for an interacting system they can be determined by the equation~\cite{LaiHaus1989B}
\begin{equation}
A({\cal P'}) = A({\cal P}) \frac{p_{{\cal P}(k+1)}-p_{{\cal P}(k)}-i/(N-1)}{p_{{\cal P}(k+1)}-p_{{\cal P}(k)}+i/(N-1)},                                                                                                                                                                                                                                                                                                                                                                                                                                                                                                                                                                                                                                                                                            \end{equation}
where ${\cal P'}$ is the permutation swapping the $k$th and $(k+1)$th indices, the coefficient of the identity permutation is determined by the normalization condition.
The center-of-mass motion is independent of the interactions, and so will have eigenstates of plane waves.  In the case of attractive interactions.  These momenta can also have very specific imaginary components corresponding to bound clusters of atoms. The ground state of relative motion occurs for 
\begin{equation}
 p_k = i \frac{N+1-2k}{2(N-1)} 
\end{equation}
 in which all the permutation coefficients apart from one (the identity permutation) are equal to zero.  Higher eigenstates can have multiple bound state clusters or strings, each with an associated real momentum $P_m$ and imaginary components (that must sum to zero) which are spaced in units of $1/(N-1)$. If we have $\eta$ clusters, each of size $n_m$, we have momenta associated with the $m$th cluster given by
\begin{equation}
 p_k = P_m +  i \frac{\sum_{\ell=1}^m n_\ell +1-2k}{2(N-1)} \; ,  \quad \sum_{\ell=1}^{m-1} n_\ell < k \le \sum_{\ell=1}^{m} n_\ell \; .
\end{equation}
This state would normally be denoted as $\ket{n_1, p_1,n_2,p_2,\ldots,n_{\eta},p_{\eta}}$, the total energy of the state, $E = \sum_k p_k^2/2$, scales as though these are $\eta$ isolated single soliton states and thus the energy eigenvalue is 
\begin{equation}
 E = \sum_{m=1}^M \left(\frac{n_m P_m^2}{2}-\frac{n_m(n_m^2-1)}{24(N-1)^2} \right) \; .
\end{equation}
The total number of different combinations of clusters scales $p(N)$, the number of ways to partition $N$ with integers (cf. Appendix \ref{sec:basisred} and \cite{AbramowitzStegunBook1984}) 

\section{Normalization, energy and overlap using the variational state}

\subsection{Preamble}

In order to make use of our variational state given in \refeq{eq:varpsifn}, we must calculate the normalization constant and expectation value of energy.  
Calculations for the energy and normalization constants for all the eigenstates in free space ($\gamma = 0$) can readily be found in literature \cite{CalabreseCaux2007}, it is also the case that the center-of-mass component of the Hamiltonian can be considered separately and so taking a finite center-of-mass component does not significantly alter the calculations.  However, the choice of normalization condition for a non local system is somewhat arbitrary and conventions vary between papers, we choose a normalization that means both the relative and center-of-mass parts are normalized to unity with respect to Jacobi coordinates.  Also most derivations of the energy rely on the fact that the gradient discontinuity at the points $x_k = x_j$ in the wave-function, exactly cancel the interaction terms (essentially from the condition of being an eigenstate), and thus these terms can simply be ignored.  Because of our variation of $\lambda$, this will no longer be the case and thus we are forced to make a more explicit calculation of the kinetic energy. In addition to this we derive a first order energy correction to the relative degrees of freedom, which is a new result.  

\subsection{Normalization \label{sec:norm}}

In order to normalize \refeq{Eq:FreeBoundState}, we could insist that two states with different center-of-mass momenta are orthonormal, i.e.~$\bket{p',N}{p,N} = \delta(p'-p)$ such as was calculated in~\cite{LaiHaus1989B} or consider wave-function to be trapped in a box which we allow to grow infinitely large~\cite{Castin2009}.  However we are interested in the normalization of the free space solution with a Gaussian center-of-mass envelope and freedom to tune a variational parameter, denoted  $\psi_{\rm var}^{(\lambda)}$ in \refeq{eq:varpsifn}.  This result and technique will also be used in appendix \ref{E1calc} and follows the method of~\cite{LaiHaus1989B}.  We consider a Fourier decomposition of the wave-function
\begin{align}
 \bket{x_1,..,x_N}{\psi_{\rm var}^{(\lambda)}} = &{\cal N}_{\lambda} \exp\left(-\sum_{k=2}^{N} \sum_{j=1}^{k-1} \frac{\sigma}{2} \vert x_{k} -x_{j} \vert\right) \nonumber \\
 &\exp\left(- \frac{N \gamma}{2} \left[\sum_{k} \frac{x_{k}}{N} \right]^2 \right)  \nonumber \\ 
= &{\cal N}_{\lambda} \int_{-\infty}^{\infty} dp \; \frac{\exp(-p^2/2\gamma)}{\sqrt{2 \pi\gamma}}  \nonumber \\
&\times \exp\left(ip\sum_{k} \frac{x_{k}}{\sqrt{N}} - \frac{\sigma}{2}\sum_{k=2}^{N} \sum_{j=1}^{k-1} \vert x_{k} -x_{j} \vert\right) \; ,
\label{eq:fourierdecomp}
\end{align}
with $\sigma = 1/(N-1)$ corresponding to \refeq{Eq:FreeBoundState}, however for greater generality we allow this parameter to be free in order to use these results for variational calculations where $\sigma \to \lambda/(N-1)$, which would correspond to \refeq{eq:varpsifn}.  Also one may wish to consider instead units in which the harmonic oscillator frequency and length are set to unity and the interaction constant rescaled to $\tilde{g}$, in which case the replacement $\sigma \to \lambda \vert \tilde{g} \vert$ would be used instead, or indeed in S.I. units $\sigma \to \lambda M\vert g_{1d}\vert /\hbar^2$.  Essentially this term serves to allow easy conversion between unit systems and making variational manipulation easier.

In the form of \refeq{eq:fourierdecomp}, it is far simpler to perform the integrals of the coordinate variables.  Calculating $\bket{\psi_{\rm var}^{(\lambda)}}{\psi_{\rm var}^{(\lambda)}}$ in coordinate space will require integration over $N$ spatial integrals and two momentum integrals, however we only need to integrate over the simplex region $x_1 \le x_2 \ldots \le x_N$ as by Bose symmetry any integration over any such region will be identical, hence we multiply by factor of $N!$ to include all possibilities for such a regions construction.  Within this simplex region, all arguments in the absolute value signs are positive and the wave-function is given by 
\begin{align}
 \bket{x_1,..,x_N}{\psi_{\rm var}^{(\lambda)}} = &{\cal N}_{\lambda} \int_{-\infty}^{\infty} dp \; \frac{\exp(-p^2/2\gamma)}{\sqrt{2 \pi\gamma}}  \nonumber \\
&\exp\left(\sum_{k} \frac{ipx_{k}}{\sqrt{N}}+ \frac{\beta(k)x_{k} }{2} \right) \; ,
\end{align}
with $\beta(k) = (N+1-2k)\sigma$, and using the notation $\int_{-\infty \le x_1<x_2<\ldots<x_{N} \le\infty} \equiv \int_{-\infty}^{x_2} dx_1 \int_{-\infty}^{x_3} dx_2\ldots\int_{-\infty}^{\infty} dx_N$, we can now express the inner product as 
\begin{align}
 \bket{\psi_{\rm var}^{(\lambda)}}{\psi_{\rm var}^{(\lambda)}} &= N!{\cal N}_{\lambda}^2 \iint_{-\infty}^{\infty} dp_1 dp_2 \; \frac{\exp(-(p_1^2+p_2^2)/2\gamma)}{2 \pi\gamma} \nonumber \\
&\int_{-\infty \le x_1<\ldots<x_{N} \le\infty} \exp\left(\sum_{k} \frac{i(p_1-p_2)x_{k}}{\sqrt{N}} + \beta(k) x_{k}\right) \; ,
\end{align}
transformation of variables $p = (p_1+p_2)/2$ and $p' = p_1-p_2$ with Jacobian unity then allows us to perform the integral over $p$ leaving
\begin{align}
 \bket{\psi_{\rm var}^{(\lambda)}}{\psi_{\rm var}^{(\lambda)}} &= N!{\cal N}_{\lambda}^2 \int_{-\infty}^{\infty} dp' \; \frac{\exp(-p'^2/4\gamma)}{2 \sqrt{\pi \gamma}}\nonumber \\
&\int_{-\infty \le x_1<\dots<x_{N} \le\infty} \exp\left(\sum_{k} ip'x_{k}/\sqrt{N} + \beta(k) x_{k}\right) \; .
\label{startpoint}
\end{align}
To perform the remaining integrals we note that $\int_{-\infty}^y dx \exp(ax+by) = \exp[(a+b)y]/a$, denoting
\begin{equation}
a(k) = \sum_{l=1}^k \beta(l) = \sigma k(k-N) \; , 
\label{eq:a(k)}
\end{equation}
 and noting $a(N)=0$, we can recursively use the previous result to perform all but one of the spatial integrals and obtain
\begin{align}
 \bket{\psi_{\rm var}^{(\lambda)}}{\psi_{\rm var}^{(\lambda)}} = N!{\cal N}_{\lambda}^2 \int_{-\infty}^{\infty} dp' \; &\frac{\exp(-p'^2/4\gamma)  A(N,p')}{2 \sqrt{\pi \gamma}}  \nonumber \\
&\int dx_N \exp\left(ip'x_N\sqrt{N} \right) \; ,
\end{align}
with 
\begin{equation}
 A(\ell,p') = \prod_{k=1}^{\ell-1} \left[ a(k)+\frac{ip'k}{\sqrt{N}} \right]^{-1} \; .
\label{eq:A(l,p)def}
\end{equation}
Integrating the final term gives $2\pi\delta(p'\sqrt{N})$ and the momentum integral is then trivial, noting that $A(N,0) = 1/(N-1)!^2\sigma^{(N-1)}$ gives us the final result for the normalization factor
\begin{align}
 {\cal N}_{\lambda}  &=  \sqrt{\sqrt{\frac{\gamma}{ N\pi}}(N-1)!\sigma^{(N-1)}} \; .  
\label{eq:normwithenvelope}
\end{align}
As was mentioned before, the normalization factor for \refeq{eq:varpsifn} in our units is obtained by letting $\sigma = \lambda/(N-1)$, in the case of $\lambda = 1$ where this relates the ground state in infinitesimal trapping \refeq{Eq:FreeBoundState}, we refer to this constant simply as ${\cal N}$. It is also worth noting that both the com wave-function and the relative are both chosen to be normalized to unity with respect to Jacobi coordinates, hence the $(\gamma/N\pi)^{1/4}$ relates to the center-of-mass part and the rest to the relative component.

\subsection{Kinetic and interaction energy \label{sec:varchanges}}

We wish calculate the kinetic energy and potential energy of the variational state, i.e. the expectation of \refeq{Eq:ScaledFirstQuantizedH} with $\lambda$ set to zero on \refeq{eq:varpsifn}, which we will denote $\hat{\rm H}_{\rm free}$.  Due to the separability of the wave-function and Hamiltonian, it is sufficient to consider only the relative part of the wave-function and note that the center-of-mass kinetic energy is given by $\gamma/4$.  We first denote $\varphi(x_1,..,x_N) = \exp\left( -\sigma\sum_{k=2}^{N} \sum_{j=1}^{k-1}\vert x_k - x_j \vert /2 \right)$, being the relative part of the variational wave-function (up to a normalization factor) and calculate the second derivative with respect to some coordinate $x_\ell$ 
\begin{align}
 &-\frac{1}{2} \pdsq{x_\ell} \varphi(x_1,..,x_N) \nonumber \\
 = &\frac{\sigma}{4}\left[ -\frac{\sigma}{2} \left( \sum_{k \ne \ell} \pd{x_\ell} \vert x_\ell - x_k \vert \right)^2 +\left( \sum_{k \ne \ell} \pdsq{x_\ell} \vert x_\ell - x_k \vert \right)\right]  \nonumber \\ &\times \varphi(x_1,..,x_N)  \nonumber \\
= &\frac{\sigma}{4}\left[ -\frac{\sigma}{2} \left( \sum_{k \ne \ell} \text{sgn} (x_\ell-x_k) \right)^2 + 2\left( \sum_{k \ne \ell} \delta (x_\ell-x_k) \right)\right] \nonumber \\ &\times \varphi(x_1,..,x_N)  \; .
\label{eq:kineticEvar}
\end{align}
The first term in \refeq{eq:kineticEvar} can be split up into terms of the form $\text{sgn}^2(x_\ell-x_b)=1$, of which there are $(N-1)$ and terms of the form $\text{sgn}(x_\ell-x_a)\text{sgn}(x_\ell-x_b)$ with $a\ne b$, of which there are $(N-1)(N-2)$.  The former will evaluate to unity by normalization of the wave-function, however the latter terms will equal $+1$ when $x_\ell < x_a<x_b$ or $x_\ell < x_b<x_a$ and when $x_a<x_b < x_\ell$ or $x_b<x_a < x_\ell$ and $-1$ when $x_a <x_\ell <x_b$ or $x_b <x_\ell <x_b$;  the wave-function must be identical in all these 6 simplicies due to Bose symmetry and so the expected value of these terms will equal $1/3$.  When the sum over all $\ell$ is performed, these terms will total to  $N(N-1)\sigma^2(1+(N-2)/3)/4$. 
The latter terms of the form $\delta(x_\ell-x_k)$ can then be combined with those from the interaction part of the Hamiltonian, noting that there are twice as many terms but $\delta(a-b) = \delta(b-a)$.  Reinstating $\sigma = \lambda/(N-1)$ we have
\begin{align}
&\bra{\psi_{\rm var}^{(\lambda)}} -\frac{1}{2}\sum_{k=1}^N \pdsq{x_k} - \frac{1}{N-1}\sum_{k=2}^{N} \sum_{j=1}^{k-1} \delta(x_k-x_j) \; \ket{\psi_{\rm var}^{(\lambda)}}  \nonumber \\
= &-\frac{\lambda^2 N}{8(N-1)}\left(1 + \frac{N-2}{3}\right)^2  + \frac{\gamma}{4}   \nonumber \\
&+  \frac{\lambda - 1}{N-1} \left[  \bra{\psi_{\rm var}^{(\lambda)}} \sum_{k=2}^{N} \sum_{j=1}^{k-1} \delta(x_k-x_j) \; \ket{\psi_{\rm var}^{(\lambda)}} \right] \;.
\label{eq:keplusint}
\end{align}
all that remains now to calculate the value of the expectation value of the delta function terms.  Following the method in Appendix \ref{sec:norm} we integrate over a simplex region $-\infty < x_1 <x_2 ... < x_N < \infty$, as a result of this we need only consider the $N-1$ terms of the form $\delta(x_k-x_{k+1})$ as the rest will be zero.  Each integral will be the same as in Appendix \ref{sec:norm} except missing a factor of $2/a(k)$ for each term $\delta(x_k-x_{k+1})$, hence the result will equal $\sum_{k=1}^{N-1} a(k)/2$ (using the result $\int_{-\infty}^{y} dx f(x,y) \delta(x-y) = f(y,y)/2$). Hence  
\begin{align}
\bra{\psi_{\rm var}^{(\lambda)}} \sum_{k=2}^{N} \sum_{j=1}^{k-1} \delta(x_k-x_j) \ket{\psi_{\rm var}^{(\lambda)}} &= \frac{\sigma}{2} \sum_{k=1}^{N-1} k(N-k) \nonumber \\
										    &= \frac{\lambda(N+1)N(N-1)}{12(N-1)} \;,
\end{align}
finally, substituting in this result into \refeq{eq:keplusint} we have
\begin{align}
&\bra{\psi_{\rm var}^{(\lambda)}} -\frac{1}{2}\sum_{k=1}^N \pdsq{x_k} - \frac{1}{N-1}\sum_{k=2}^{N} \sum_{j=1}^{k-1}  \delta(x_k-x_j) \ket{\psi_{\rm var}^{(\lambda)}} \nonumber \\ 
&= \frac{N(N+1)}{24(N-1)} [-\lambda^2+2\lambda(\lambda -1)] + \frac{\gamma}{4} \nonumber \\ 
&= E_0  (2\lambda - \lambda^2) + \frac{\gamma}{4}  \; ,
\end{align}
where $E_0 = -N(N+1)/24(N-1)$.  This discussion has not mentioned the harmonic envelope of the center-of-mass function, however due to the separability of the Hamiltonian this will only add factor of the center-of-mass kinetic energy, which will be independent of $\lambda$ regardless of what the center-of-mass wave-function is. This energy term is combined with the potential energy calculation derived in Appendix \ref{E1calc} to form the basis for a variational principle.

\subsection{Derivation of the first-order perturbation energy \label{E1calc}}
This section is related to the calculation of $\bra{\psi_{\rm var}^{(\lambda)}} V(x) \ket{\psi_{\rm var}^{(\lambda)}}$, where $\hat{V}(x) = \gamma^2 \sum_{k=1}^N x_{k}^2/2$, note that \refeq{foecor} is given by this quantity minus the center-of-mass energy. It is again easier not to perform this integral in Jacobi coordinates but to Fourier transform out the center-of-mass, we will also again replace the factor $1/(N-1)$ with $\sigma$ to generalize the results for our variational principle.  This calculation is similar to, although more complicated than, the calculation of the normalization factor; to that end we can start the calculation from~\refeq{startpoint} (as no spatial integrals are yet performed) adding in the potential factor $V(x)$ giving 
\begin{align}
 \bra{\psi_{\rm var}^{(\lambda)}} V(x) \ket{\psi_{\rm var}^{(\lambda)}} &= \frac{\sqrt{N}}{2 \pi A(N,0)} \int_{-\infty}^{\infty} dp' \; \exp\left(\frac{-p'^2}{4\gamma} \right) \nonumber \\
&\int_{-\infty \le x_1<x_2<\ldots<x_{N} \le\infty}  \frac{\gamma^2}{2} \sum_k x_{k}^2 \nonumber \\
&\times \exp\left(\sum_{k} ip'\frac{x_{k}}{\sqrt{N}} + \beta(k) x_{k}\right) \; .
\label{startpoint2}
\end{align}  
The same recursive integral procedure can be applied here except we need two additional results, true for $\text{real}(k)>0$
\begin{align}
 \int_{-\infty}^{y} dx \; x^n \exp(kx) =
\begin{cases}
\frac{1}{k} \exp(ky) & \text{if $n=0$,}
\\
\frac{ky-1}{k^2} \exp(ky) &\text{if $n=1$.}
\\
\frac{(ky)^2-2ky+2}{k^3} \exp(ky) &\text{if $n=2$.}
\end{cases}
\label{exps}
\end{align}
Let us consider only the latter part of~\refeq{startpoint2} omitting the constant $\sqrt{N}\gamma^2/4 \pi A(N,0) $, taking the integrals in order from $x_1$ to $x_N$, the integral over $x_{\ell}$ will be over a function of the form
\begin{align}
I(\ell) = &A(\ell,p') \left(k_0(\ell) +k_1(\ell)x_{\ell} + k_2(\ell)x_{\ell}^2 + \sum_{\ell'=\ell+1}^N x_{\ell'}^2\right)\nonumber \\
&\times\exp\left(a(\ell)x_{\ell'}+ \sum_{\ell'=\ell+1}^N \beta(\ell')x_{\ell'} + \frac{ip\ell x_{\ell}}{\sqrt{N}}\right)\; ,
\label{eq:lthcomp}
\end{align} 
 with $k_0(1) = k_1(1) = 0$ and $k_2(1) = 1$ and $A(\ell,p')$ defined in \refeq{eq:A(l,p)def}.   The common prefactor of $A(\ell,p')$ is the equivalent of $k$ from~\refeq{exps}.  Besides this, each integral will increase the factor in front of the $x_\ell^2$ term by one each time and hence $k_2(l) = l$.  Contributions to $k_1(\ell+1)$ come from $k_1(\ell)$ and $k_2(\ell)$ and as such~\refeq{exps} implies $k_1(\ell+1) = k_1(\ell) - 2 k_2(\ell) (a(\ell)+ip\ell/\sqrt{N})^{-1}$, given that $k_1(1) =0$ this implies 
\begin{align}
 k_1(\ell+1) &= -2 \sum_{k=1}^{\ell} \frac{k}{a(k)+\frac{ikp'}{\sqrt{N}}} \\
k_1(N)|_{p'=0} &=-\frac{2}{\sigma} \sum_{k=1}^{N-1} \frac{1}{k}\; .
\end{align}
Applying this same induction logic down to $k_0$ gives
\begin{align}
 k_0(\ell+1) &= k_0(\ell) - \frac{k_1(\ell)}{a(\ell)+\frac{ip'\ell}{\sqrt{N}}}+ 2\frac{k_2(\ell)}{(a(\ell)+\frac{ip'\ell}{\sqrt{N}})^2} \nonumber \\ 
	     &=  \sum_{\ell'=1}^{\ell} \sum_{k=1}^{\ell'} \frac{2 k}{(a(k)+\frac{ikp'}{\sqrt{N}})(a(\ell')+\frac{i\ell'p'}{\sqrt{N}})} \nonumber \\ 
 k_0(N)|_{p'=0}    &= \frac{2}{\sigma^2} \sum_{\ell=1}^{N-1} \sum_{k=1}^{\ell} \frac{1}{\ell(N-\ell)(N-k)} \;,
\end{align}
simply evaluating these at $N$ and performing the final integration over $x_N$ then yields
\begin{align}
\int_{-\infty}^{\infty} dx_N \; I(N) = &\frac{ 2 A(N,p')\pi}{\sqrt{N}} \left[-\frac{k_2(N) \delta''(p')}{N} \right. \nonumber \\
  &+ \left. i \frac{k_1(N) \delta'(p')}{\sqrt{N}} + k_0(N) \delta(p')\right] \; ,
\label{eq:deltaints2}
\end{align} 
we then insert this expression back into~\refeq{startpoint2} giving
\begin{align}
 \bra{\psi_{\rm var}^{(\lambda)}} V({\bf x}) \ket{\psi_{\rm var}^{(\lambda)}} &= \frac{\gamma^2}{2 A(N,0)} \int_{-\infty}^{\infty} dp' \exp\left(\frac{-p'^2}{4\gamma} \right)  \nonumber \\
&\times\left[-\frac{k_2(N) \delta''(p')}{N} + i \frac{k_1(N) \delta'(p')}{\sqrt{N}} + k_0(N) \delta(p')\right]\; .
\end{align}
The integral over the $\delta(p')$ term can be performed immediately and gives $\gamma^2 k_0(N)/2$. Considering next the integral over $ \delta'(p')$; since $\exp(-p'^2/4\gamma)$ has zero gradient at the origin it will not contribute, however the terms
\begin{align}
\pd{p} k_1(N)|_{p'=0} &= \frac{2 i }{\sigma^2 \sqrt{N}} \sum_{k=1}^{N-1} \frac{1}{k^2} \nonumber \\
\pd{p} A(N,p')|_{p'=0} &= -\frac{i }{\sigma \sqrt{N}} A(N,0)  \sum_{k=1}^{N-1} \frac{1}{k} \; ,
\end{align} 
will contribute to \refeq{eq:deltaints2}, giving
\begin{align}
 &\int_{-\infty}^{\infty} \; dp' i\delta'(p') \exp\left(\frac{-p'^2}{4 \gamma}\right) A(\ell,p') \frac{K_1(N)}{\sqrt{N}} \nonumber \\
&=  -2 A(N,0) \frac{1}{N \sigma^2} \left(\left[\sum_{k=1}^{N-1} \frac{1}{k}\right]^2+\sum_{k=1}^{N-1} \frac{1}{k^2} \right)\;.
\end{align}
Finally for the $ \delta''(p')$ term we must include $A''(N,p')\vert_{p'=0} = -A(N,0)\sum_{k,\ell=1}^{N-1} (1+\delta_{kl}) k\ell/ N a(k)a(\ell)$ and the differential of a Gaussian, hence we have
\begin{align}
 &-\int_{-\infty}^{\infty} \; dp' \delta''(p') \exp\left(\frac{-p'^2}{4 \gamma}\right) A(l,p') \nonumber \\
&=  A(N,0) \left(\frac{1}{2 \gamma} + \frac{1}{N \sigma^2} \left\{ \left[\sum_{k=1}^{N-1} \frac{1}{k} \right]^2 +\sum_{k=1}^{N-1} \frac{1}{k^2} \right\} \right)\;.
\end{align}
Summing these three terms together, and substituting $k_2(N)=N$, we are left with
\begin{align}
 \bra{\psi_{\rm var}^{(\lambda)}} V \ket{\psi_{\rm var}^{(\lambda)}} =&  \frac{\gamma^2}{\sigma^2} \left[ \frac{\sigma^2}{4\gamma}  +  \sum_{\ell=1}^{N-1} \frac{1}{\ell(N-\ell)} \sum_{k=1}^{\ell} \frac{1}{N-k}  \right. \nonumber \\
&\left. - \frac{1}{2N} \left(\left[\sum_{k=1}^{N-1} \frac{1}{k} \right]^2 + \sum_{k=1}^{N-1} \frac{1}{k^2} \right) \right] \; .
\label{eq:e1pluscom}
\end{align}
The first term in this expression is equal to $\gamma/4$, which is simply the potential energy of the center-of-mass component.
 It can be proved via induction~\cite{BillamPrivate2011} that the double sum is equal to
\begin{align}
 \sum_{\ell=1}^{N-1} \frac{1}{\ell(N-\ell)} \sum_{k=1}^{\ell} \frac{1}{N-k}  =\frac{1}{2N} \left[ \left(\sum_{k=1}^{N-1} \frac{1}{k} \right)^2 + \sum_{k=1}^{N-1} \frac{3}{k^2} \right]  \;,
\end{align}
thus reinstating $\sigma = \lambda/(N-1)$, the remaining terms simplified down to
\begin{align}
\bra{\psi_{\rm var}^{(\lambda)}} V \ket{\psi_{\rm var}^{(\lambda)}} = \frac{\gamma^2 (N-1)^2 }{N \lambda^2} \sum_{k=1}^{N-1} \frac{1}{k^2} + \frac{\gamma}{4} \;,
\end{align}
which is used in Sec.~\ref{sec:pert and var}.  Equation \refeq{foecor} is the first order energy correction to the free soliton with Gaussian center-of-mass envelope and is obtained by subtracting the center-of-mass energy and setting $\lambda =1$
\begin{align}
E^{(1)} = \frac{\gamma^2 (N-1)^2 }{N} \sum_{k=1}^{N-1} \frac{1}{k^2} \;.
\end{align}

\subsubsection{Energy correction from potentials of higher powers of $x$}

An energy correction for general power law potentials can be derived in the mean field case.  For ${\rm Re}(m) > -1$
\begin{equation}
  \int_{-\infty}^{\infty} d x \; \frac{\sech(x/2)^2}{4} \frac{\vert x \vert^{m}}{2} =
\begin{cases}
 m!  \zeta(m) (1-2^{1-m}) &\quad m \ne 1\;, \\
 \log(2)   &\quad m = 1 \;,
\end{cases}
\label{xtothemsech}
\end{equation}
with $\zeta(m)$ the Riemann zeta function.  A similar result would be desirable to calculate energy correction from an anharmonic potential for a quantum soliton, although potentials with $m \ne 2,0$ will couple the center-of-mass and relative degrees of freedom together (possibly only very weakly) and so $[\hat{H}_{\rm cm},\hat{H}_{\rm rel}] \ne 0$ and this is only of limited use.

\subsection{Overlap of the relative components of the variational wavefunctions \label{sec:varchanges2}}
Finally we consider the overlap between the relative parts of the variational wavefunction with $\lambda > 1$ and the ground state in infinitesimal trapping $\gamma=0, \; \lambda = 1$, given by
 \begin{align}
 \bket{\psi_{\rm var}^{(\lambda)}}{\psi_{\rm var}^{(1)}} = &{\cal N}_1 {\cal N}_{\lambda}  \int_{-\infty}^{\infty} dx_1 \ldots\int_{-\infty}^{\infty} dx_N  \vert \psi_{\rm cm} \vert^2  \nonumber \\
&\times  \exp\left( -\frac{\lambda+1}{2(N-1)} \sum_{k=2}^{N} \sum_{j=1}^{k-1} \vert x_{k} -x_{j} \vert \right)  \; .
 \end{align}
This calculation can be achieved by performing the calculations in Appendix \ref{sec:norm} with $\sigma \to (1+\lambda)/2(N-1)$, the resulting factor will not equal unity and instead will be equal to ${\cal N}_{1} {\cal N}_{\lambda}/ {\cal N}_{(1+\lambda)/2}^2$.  Therefore that the overlap is given by
 \begin{align}
\bket{\psi_{\rm var}^{(\lambda)}}{\psi_{\rm var}^{(1)}}  &=  \frac{\lambda^{(N-1)/2}}{\prod_{k=1}^{N-1} (1+\lambda)/2} \nonumber \\
&= \left(\frac{2 \sqrt{\lambda}}{(1+\lambda)}\right)^{N-1} \; ,
 \end{align}
which is used in Sec \ref{sec:varminim}. 

\section{Ground state energy for $H_{\textrm{R}}(\xi_{2})$ in the interaction dominated regime\label{App:GroundStateEnergy}}

Using the identity \cite{WolframGammaFunction}
\begin{equation}
\frac{\Gamma(z+ 1/2)}{\Gamma(z)} =
\sqrt{z}\left(
1 - \frac{1}{8z} 
+ \sum_{k=2}^{\infty}\frac{c_{k}}{z^{k}}
+ \cdots
\right),
\end{equation}
where the $c_{k}$ are coefficients for the higher order terms in the asymptotic expansion, we see from \refeq{Eq:GammaFunctionIdentity} that
\begin{equation}
\sqrt{-\nu_{0}}\left[
1 + \frac{1}{8\nu_{0}} 
+ \sum_{k=2}^{\infty}\frac{c_{k}}{(-\nu_{0})^{k}}
+ \cdots
\right]
=
\frac{1}{2\sqrt{2 \gamma}}.
\label{Eq:RootNuZeroTruncated}
\end{equation}

Hence, taking the limit $\gamma \to 0$ (interaction dominated regime) implies $\nu_{0}\to -\infty$ , and we may truncate the asymptotic series.  To lowest order $\sqrt{-\nu_{0}} \approx 1/2\sqrt{2\gamma}$, which we substitute into the right hand side of [rearranged from \refeq{Eq:RootNuZeroTruncated}]
\begin{equation}
\sqrt{-\nu_{0}} = 
\frac{1}{2\sqrt{2\gamma}}
+
\frac{1}{8\sqrt{-\nu_{0}}} + \mathcal{O}(\nu_{0}^{-3/2}),
\end{equation}
squaring the result to get
\begin{equation}
\nu_{0}= 
-\frac{1}{\gamma}\left[
\frac{1}{8} + \frac{\gamma}{4} + \mathcal{O}(\gamma^{2})
\right].
\label{Eq:NuZeroTruncated}
\end{equation}
Hence, substituting \refeq{Eq:NuZeroTruncated} into \refeq{Eq:TwoBodyEnergy} for $n=0$ yields
\begin{equation}
\lim_{\gamma\to 0}E_{\textrm{R},0} = -\frac{1}{4} + \mathcal{O}(\gamma^{2}).
\end{equation}

\section{Energy correction to the Hartree product state \label{comke}}  

This section derives the energy correction to the Hartree product state $\ket{\Psi_{\rm H}}$, this state is a product of $N$ identical single particle wavefunction $\Phi(x) = \sech(x/2)/2$.  The first result we require is the potential energy correction to each single particle wavefunction
\begin{align}
\frac{1}{2} \int_{-\infty}^{\infty} \vert \Phi(x) \vert^2 \gamma^2 x^2  &= -\frac{\gamma^2}{8} \int_{-\infty}^{\infty} d x \; x^2 \sech^2(x/2) \nonumber \\
               & = \frac{\gamma^2\pi^2}{6} \; ,
\end{align}
the total energy correction is thus $N$ times this value.  However we are interested only in the relative energy correction given by 
\begin{align}
 E^{(1)}_{\rm H} &= \gamma^2\bra{\psi_{\rm H}} \frac{1}{2} \sum_{k=1}^N x_k^2 - \frac{1}{2} \left(\sum_{k=1}^{N} \frac{x_{k}}{N}\right)^2 \ket{\psi_{\rm H}}\nonumber \\ 
		 &=  \gamma^2\bra{\psi_{\rm H}}  \frac{N-1}{2N} \sum_{k=1}^N x_k^2 - \sum_{k<j} x_k x_j \ket{\psi_{\rm H}}\;.
\end{align}
All the cross terms of the form $x_k x_j$ will evaluate to zero as $\sech(x)$ is an even function, thus leaving only the power terms. By Bose symmetry $\ev{f(x_k)} = \ev{f(x_j)}$ and thus the value of all the terms in the first sum will be identical to the single particle correction and we have
\begin{align}
 E^{(1)}_{\rm H} = (N-1) \frac{\gamma^2\pi^2}{6} \;.
\end{align}

\section{Exact solution to the variational minimization \label{sec:quartic}}

The solution derived to the minimization equation~(\ref{eq:varminpoly}) is given by
\begin{equation}
 \lambda^3(\lambda-1) -\kappa = 0 \;,
\end{equation}
with $\kappa>0$ defined by \refeq{eq:kdef}. 
This equation has exactly one real positive solution $\lambda_0$ corresponding to an energy minimum, this solution can be derived analytically~\cite{Mathematica7010} (cf.\ Ref.~\cite{BillamWrathmall2011b}); it is given by:
\begin{align}
\label{eq:lambda0exact}
\lambda_0&\equiv\frac{1}{4}\left(1+ \sqrt{\Lambda} + \sqrt{3-\Lambda + 2 \Lambda^{-1/2}}\right)\;,
\end{align}
with
\begin{align}
\Lambda &=1-\frac{16 \left(\frac{2}{3}\right)^{1/3} \kappa}{Y}+2 \left(\frac{2}{3}\right)^{2/3} Y \;, \\
Y &=\left(-9 \kappa+\sqrt{3} \sqrt{27 \kappa^2+256 \kappa^3}\right)^{1/3} \;.
\end{align}
A Taylor expansion about $\kappa=0$ yields 
\begin{equation}
\lambda_0 = 1+\kappa-3 \kappa^2+O(\kappa^{3}).
\end{equation}

\section{Truncating the Hilbert space by introducing energy cut-offs and projection to the zero center-of-mass excitation subspace \label{truncate}}

\subsection{Integer partition and energy level degeneracy \label{subsec:intpart}}

The relation between energy level degeneracy in systems of identical particles and number partitioning has been investigated in ~\cite{SchumayerHutchinson2011,WeissHolthaus2002,KubasiakKorbicz2005} and references therein. Within a one dimensional Harmonic oscillator, the degeneracy for $N$ distinguishable particles scales the same as the degeneracy for one particle in an $N$ dimensional spherically symmetric potential.  This is not the case for indistinguishable particles, to calculate these we must use introduce integer partition functions.
We introduce the notation $p([a,b],m)$ being the number of ways to partition an integer $m$ using only integers $a\le z \le b$, in order to compute these for a given $b$, we use the recursion relation 
\begin{align}
p([a,b],m) =
\begin{cases} &0 \qquad \text{if $\; a \ge \rm{min}(m,b)$ and $m \ne 0$} 
\\
 &1 \qquad  \text{if $\; [a = m \; \text{ or } \; m = 0 ]\: \& \: a \le b$}
\\ &p([a+1,b], m) + p([a,b], m - a) \; \text{otherwise} \; .
\end{cases}
\label{eq:recursion}
\end{align}
This works by noting that we can divide a partition into two distinct sets, partitions which uses only numbers larger than $a$, being $p([a+1,b], m)$, and partitions which uses $a$ at least once in the partitions, $p([a,b], m - a)$.   Also $p([a,b], 0) =0$ by convention.

Using the usual Fock space representation of these harmonic oscillator states $\ket{N_0,N_1,\ldots }$ with $\sum_k N_k = N$ and defining $\tilde{E}$ as the energy of the state (with no interactions) minus the ground state energy divided by $\gamma$
\begin{equation}
 \tilde{E} = \frac{E}{\gamma}  - \frac{N}{2} =  \sum_{k=0}^{\infty} k N_k \; .
\end{equation}
Given that each occupancy of the $k$th mode raises the energy by $k$ it can be seen that the degeneracy of the energy level $\tilde{E}$ is given by the number of ways to partition $\tilde{E}$ using $N$ nonnegative integers.  
Denoting $\Phi(\tilde{E},\ell)$ as the ways to partition $\tilde{E}$ in $\ell$ numbers we have
\begin{equation}
 g(\tilde{E},N) = \sum_{\ell=0}^N \Phi(\tilde{E},\ell) \;.
\end{equation}
It is also known that this sum is equal to the number of ways to partition an integer `$\tilde{E}$' using only numbers less than or equal to $N$ i.e. $g(\tilde{E},N) = p([1,N], \tilde{E})$. 

\subsection{Truncation with an energy cut-off}

In order to make a basis computationally manageable, it must truncated to be made to be finite.  This is achieved by only taking states with energy less than an arbitrary cut-off $\scut$, note that this also implies that $N_k = 0$ if $k>\scut$.  The size of this truncated Fock state basis is given by
\begin{equation}
 \sum_{\tilde{E}=0}^{\scut} g(\tilde{E},N) = \sum_{\tilde{E}=0}^{\scut} p([1,N], \tilde{E})\;.
\end{equation}
The reason an energy cut-off is chosen rather than a mode cut-off at $\scut$ (although as mentioned before this is implicit in an energy cut-off method) is two fold.  Firstly in order to project into the center-of-mass and relative excitation basis we require all the states with a given energy $\tilde{E}$ [the Hamiltonian (\ref{eq:comham}) is block diagonal], if we do not have all those states the projection is not possible.  Secondly having just a mode cut-off would include the state $\ket{0,0,\ldots,N }$ with $\tilde{E}=N\scut$, but not the state $\ket{N-1,0,\ldots0,1 }$ (one occupancy in the $\scut+1$th mode) with $\tilde{E} = \scut+1$, as long as harmonic oscillator energy remains a non negligible quantity, the former state will have almost no mixing to the ground state, making it a very inefficient truncation. 

\subsection{Deriving the projector to the center-of-mass basis \label{sec:com}}

As we have expressed the Hamiltonian (\ref{eq:compham}) in terms of $\hat{a}^{\dagger}_k$ and $\hat{a}^{\phantom{\dagger}}_k$, the creation and annihilation operator for bosons in mode $k$,  it is far simpler to compute the matrix elements in terms of basis states in the $\ket{N_0,N_1,\ldots }$, occupation notation.  Therefore we wish to calculate the elements and then project into eigenstates of the center-of-mass Hamiltonian $\hat{H}_{\rm cm}$ given in \refeq{eq:comham}.
It is therefore sufficient to diagonalize $\hat{A}^{+}\hat{A}^{-}$, [given by \refeq{eq:modecreation}] as this is the only operator dependence in $\hat{H}_{\rm cm}$,  using basis states of the form $\ket{N_0,N_1,\ldots }$.  This gives a square matrix $\hat{P}$ of eigenvectors of center-of-mass, which can project the truncated Fock state basis into this new basis, and a vector of eigenvalues.  This is computationally simple as $\hat{A}^{+}\hat{A}^{-}$ cannot mix states of different energies and therefore is block diagonal when states are ordered by energy and each block can be diagonalized separately. By removing all the columns of $\hat{P}$ with associated eigenvalues not equal to zero (meaning they have excitations in the center-of-mass mode) we are left with a rectangular matrix $\tilde{P}$ which projects into this ground state of center-of-mass excitation subspace that we call the `reduced basis'. 

Using $\tilde{P}$ results in a far smaller basis set (discussed in the next subsection) without changing any of the relative dynamics, however it is not immediately clear what states in this new basis relate to.  Given that each partition of $\tilde{E}$ into $N$ positive integers has the interpretation that each integer $k$ represents a single occupancy in the $k$th mode, one may ask what the relation to quantum numbers is of partitions in terms of integers less than or equal to $N$, for instance $\tilde{E}=2$ can be partitioned by $1+1$ and $2$.  Given that we know the ladder operator associated with the center-of-mass mode $\hat{A}^{\pm}$ , satisfies $[\hat{H}_0 ,  \hat{A}^{\pm}] = \pm \gamma \hat{A}^{\pm}$ and is thus spaced in steps of unity times $\gamma$, we can associate all the 1's in a given partition with a quanta in this mode.  Assuming we have $\ell$ quanta in the center-of-mass mode, this leaves all the numbers $2 \le z \le N$ as ways to partition $\tilde{E} - \ell$, which must then relate to some relative excitation modes.  Going back to $\tilde{E}=2$ the partition, $2=1+1$ is two quanta in the center-of-mass mode i.e. $\hat{A}^{+} \hat{A}^{+} \ket{N,0,\ldots }$ and the partition $2=2$ is one quanta in the first relative mode.  

In order to help understand this we examine the $N=2$ case in first quantization, using Jacobi coordinates [\refeq{Eq:JacobiCoordinates}].  The Hamiltonian can be expressed in two commuting parts
\begin{align}
 H_{\rm cm} &= -\frac{1}{4} \pdsq{ x_{\rm C}} + x_{\rm C}^2, \quad \rm{and}\\
H_{\rm rel} &= - \pdsq{\xi_2} + \xi_2^2/4 \;.
\end{align}
For distinguishable atoms, these would each have normal harmonic oscillator eigenstates (up to a scaling factor), which can be multiplied together to create a many-body eigenstate. However, we require Bose symmetry of the many-body wavefunction: $\psi(x_1,x_2) = \psi(x_2,x_1)$; in terms of Jacobi coordinates this implies no conditions on $ x_{\rm C}$ but that $\psi( x_{\rm C},\xi_2) = \psi( x_{\rm C},-\xi_2)$ and hence odd eigenstates for $H_{\rm rel}$ are disallowed and relative energy levels are spaced in units of 2.

\subsection{Basis size reduction \label{sec:basisred}}

As mentioned in Appendix \ref{sec:com}, the center-of-mass mode ladder operator $\hat{A}^{\pm}$ of \refeq{eq:modecreation} has an energy spacing of unity, implying relative excitation modes are spaced in units of $2,3,..,N$.  
 
In our reduced basis, the subset with the center-of-mass in the ground state, we can no longer partition $\tilde{E}$ using the number 1. Therefore the energy degeneracy $\tilde{g}(\tilde{E})$ of level $\tilde{E}$ in the reduced basis, is the number of ways to partition $\tilde{E}$ using integers $z$ satisfying $2 \le z \le N$, i.e.~$\tilde{g}(\tilde{E},N) = p([2,N],\tilde{E})$. Therefore the number of basis states in the reduced basis relative to the occupation number basis with cut of $\tilde{E}$ is given by
\begin{align}
\Delta(\scut,N) = \frac{\sum_{\tilde{E}=0}^{\scut} p([2,N],\tilde{E})}{\sum_{\tilde{E}=0}^m p([1,N],\tilde{E})} \;.
\end{align} 
We can use the equation of \refeq{eq:recursion} to write $p([2,N],\tilde{E}) = p([1,N],\tilde{E}) - p([1,N],\tilde{E}-1)$, in the sum from $0$ to $\scut$, all terms cancel apart from those at the end points of the sum, leaving only the term $p([1,N],\scut)$ and hence the size of the reduced basis is just the degeneracy of the $\scut$th energy level in the occupation number basis, thus we have 
\begin{align}
\Delta(\scut,N) = \frac{p([1,N],\scut)}{\sum_{\tilde{E}=0}^{\scut} p([1,N],\tilde{E})} \;.
\label{eq:basisreduc}
\end{align} 
Essentially this property can be seen from projecting the set of kets with energy $\scut$ into the center-of-mass excitation basis, this set will contain all the relative excited states with energy less than or equal to $\scut$, but with additional center-of-mass excitation. 

\begin{figure}[t]
\begin{center}
\includegraphics[width=0.9\linewidth]{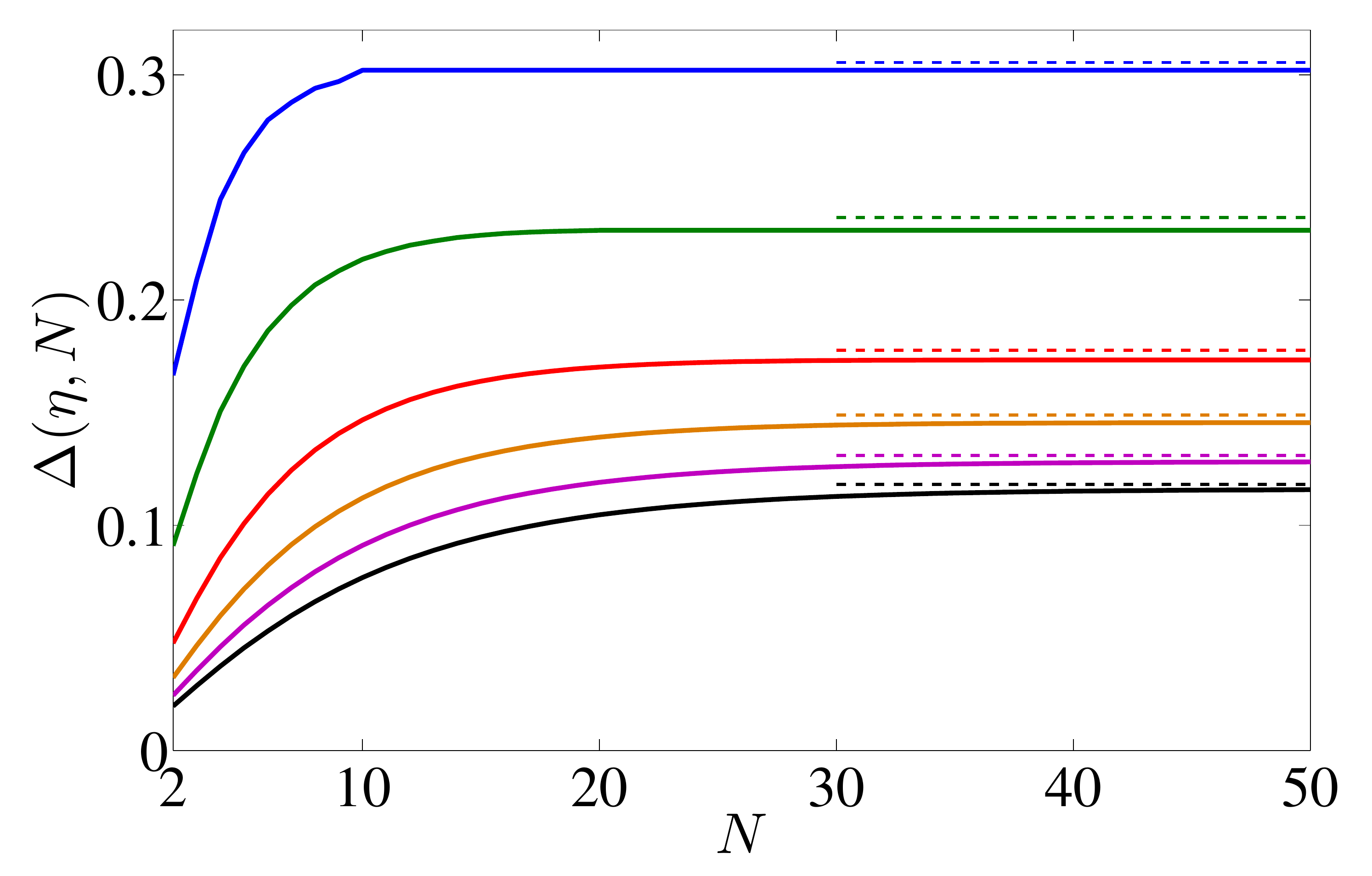}
\caption{(Color online): Reduced basis size divided by truncated basis size, given by \refeq{eq:basisreduc}, for different cut-off energies. Top to bottom lines are for cut-off energies $\scut=10,20,40,60,80,100$, dotted lines are the estimate of \refeq{eq:redest}.  Basis reduction is most significant for small $N$ but \refeq{eq:redest} provides a good upper bound on reduction for large $N$.} 
\label{fig:basisred}
\end{center}
\end{figure}

The basis reduction for $N=2$ can be calculated by noting there are $\lfloor k/2 \rfloor + 1$ ways to partition $k$ using 1 and 2 (the notation $\lfloor k \rfloor$ means round $k$ down to an integer), thus the reduced basis is $\lfloor \scut/2 \rfloor + 1$ in size, the number of states in the truncated occupation number basis is 
\begin{equation}
 \sum_{k=0}^{\scut} \left(\lfloor k/2 \rfloor + 1 \right)= 
\begin{cases} 1+\scut+\scut^2/4 & \text{if $\scut$ even}
\\
 1+\scut+(\scut^2-1)/4 &\text{if $\scut$ odd.}
\end{cases}
\end{equation}
To leading order the reduction $\Delta(\scut,N)$ goes as $2/\scut$. Such simple analytic expressions are not known for general $N$, however we have the following expression by Ramanujan~\cite{AbramowitzStegunBook1984}
\begin{equation}
p([1,N\ge \scut],\scut) \sim \frac{1}{4\scut\sqrt{3}} \exp\left(\pi \sqrt{\frac{2\scut}{3}}\right)  \; \mbox{as } \scut \to \infty,
\end{equation}
this can be used to get an asymptotic estimate of the basis reduction by replacing the sum in \refeq{eq:basisreduc} with an integral, giving
\begin{equation}
\frac{p([1,N\ge \scut],\scut) }{ \int_0^{\scut} p([1,N\ge \scut],\scut') d\scut'} \sim  \frac{\pi}{\sqrt{6\scut}} - \frac{1}{\scut} + {\cal O}(\scut^{-3/2}) \; ,
\label{eq:redest}                                                                                                                                            
\end{equation}
which will be our best estimate for the reduction achieved for large $N$,  note that this improves slower than the $\propto 1/\scut$ reduction for the $N=2$ case.  This asymptotic estimate is included in \reffig{fig:basisred}, along with the reduction for intermediate values of $N$.

\newpage

\end{appendix}

\end{document}